\documentclass[aps,twocolumn,nofootinbib,notitlepage,superscriptaddress]{revtex4-2}
\usepackage{graphicx}
\usepackage{dcolumn}
\usepackage{bm}
\usepackage{mathrsfs}

\usepackage{amsmath,amssymb}
\usepackage{float}
\usepackage{multirow}
\usepackage{slashed}
\usepackage{xcolor}
\usepackage{physics}
\usepackage{multirow}
\usepackage[colorlinks=true, pdfstartview=FitV,
bookmarks=true,bookmarksnumbered=true,breaklinks]{hyperref}
\usepackage{mathtools,braket}
\usepackage{soul}
\usepackage{lipsum}  
\usepackage{color}

\usepackage[normalem]{ulem}

\definecolor{blue}{rgb}{0.0, 0.0, 1.0}
\definecolor{red}{rgb}{1.0, 0.0, 0.0}
\definecolor{royalblue}{rgb}{0.0, 0.14, 0.4}
\hypersetup{linkcolor=royalblue, citecolor=blue, urlcolor=royalblue}

\usepackage{hyperref}
\hypersetup{colorlinks=true,citecolor=blue,linkcolor=blue,urlcolor=blue}

\usepackage[mathlines]{lineno}
\usepackage{newtxmath}
\def\orcid#1{\kern .08em\href{https://orcid.org/#1}
{\includegraphics[keepaspectratio,width=0.7em]{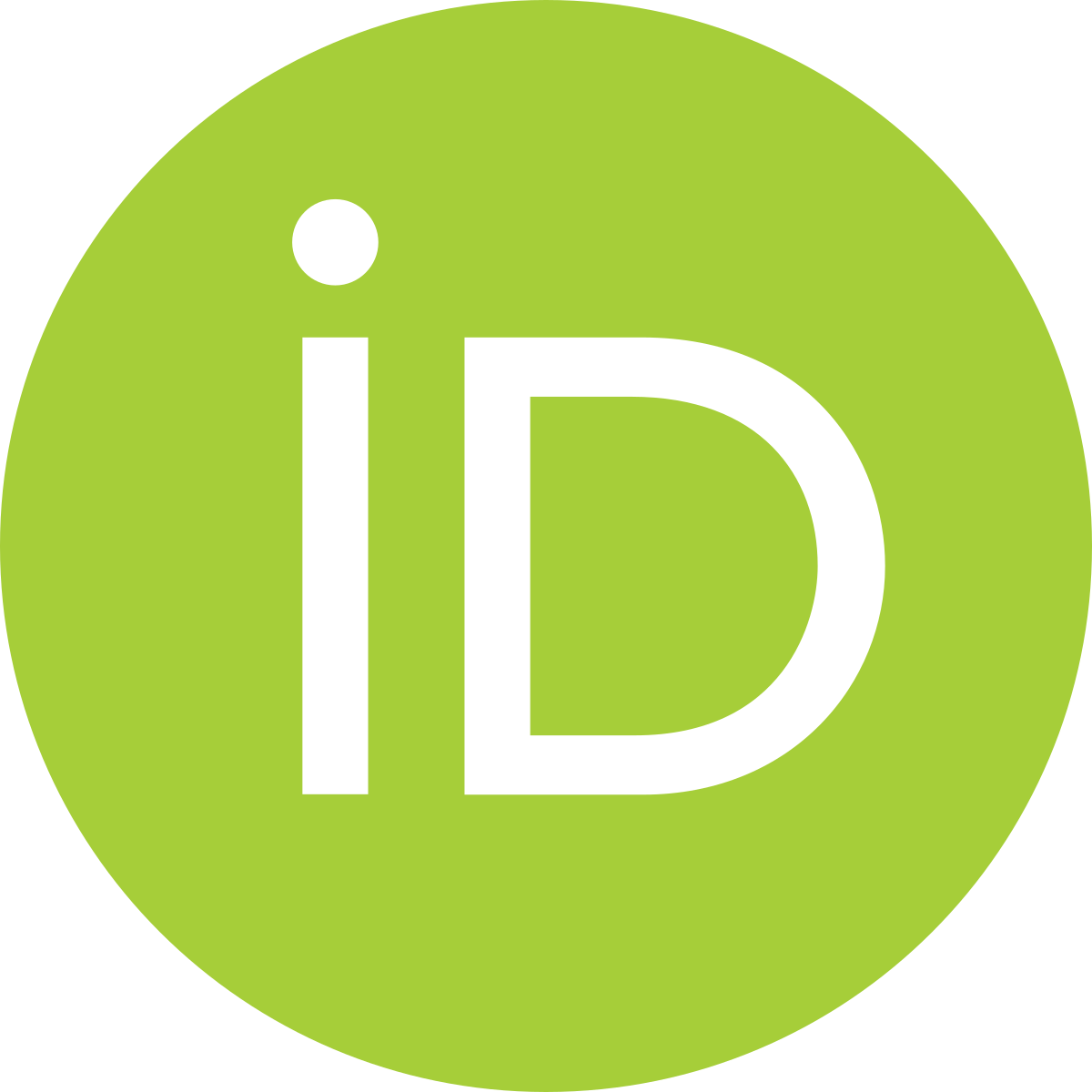}}}

\begin{document}
\vspace*{-11ex}
\begin{flushright}
{\small{\bf LFTC-23-1/74}}
\end{flushright}
\title{
\vspace{0ex} 
In-medium properties of the light and heavy-light mesons 
\\ in a light-front quark model}

\author{Ahmad Jafar Arifi\orcid{0000-0002-9530-8993}}
\email{ahmad.arifi@riken.jp}
\affiliation{Asia Pacific Center for Theoretical Physics, Pohang, Gyeongbuk 37673, Korea}
\affiliation{Few-Body Systems in Physics Laboratory, RIKEN Nishina Center, Wako 351-0198, Japan}

\author{Parada.~T.~P.~Hutauruk\orcid{0000-0002-4225-7109}}
\email{phutauruk@pknu.ac.kr}
\affiliation{Department of Physics, Pukyong National University (PKNU), Busan 48513, Korea}
\affiliation{Department of Physics Education, Daegu University, Gyeongsan 39453,
Korea}

\author{Kazuo Tsushima\orcid{0000-0003-4926-1829}}
\email{kazuo.tsushima@cruzeirodosul.edu.br; kazuo.tsushima@gmail.com}
\affiliation{Laboratório de Física Teórica e Computacional-LFTC, 
Universidade Cidade de S\~{a}o Paulo, 01506-000 S\~{a}o Paulo, SP, Brazil}
\date{\today}

\begin{abstract}
We investigate the in-medium properties of pseudoscalar and vector mesons 
with the light-light and heavy-light quarks in a light-front quark model (LFQM), 
using the in-medium quark properties computed by the quark-meson coupling (QMC) model. 
Both models are constructed on an equal footing with the constituent quark degree of
freedom.
Here, we particularly focus on the weak decay constants and distribution amplitudes (DAs) 
of the mesons in symmetric nuclear matter. 
We find that the weak decay constants decrease as nuclear density increases for $\pi$, $K$, 
$D$, and $B$ pseudoscalar as well as $\rho$, $K^{*}$, $D^{*}$, and $B^*$ vector mesons, 
where their properties in free space have good agreement with the available experimental 
and lattice QCD data. A larger reduction is found for the light-light quark pseudoscalar mesons, 
while a smaller reduction is found for the heavy-light quark vector mesons, 
in particular, with the bottom quark. 
We discuss the effect of the vector potential on the weak decay constants, 
and present our predictions for the in-medium modifications of DAs.  
Also, a comparison with the free space lattice QCD data is made.
\end{abstract}
\maketitle

\section{Introduction} \label{sec:intro}

One of the most challenging problems in hadronic and nuclear physics is 
how the properties and structure of hadron are modified in a nuclear medium, 
and how such modifications would be reflected on the observables, such as
cross-sections and extracted form factors~\cite{EuropeanMuon:1983wih,Suzuki:2002ae,Hayano:2008vn}. 
It was indicated experimentally that the nucleon structure function is
modified in nuclei, which is known as the European Muon Collaboration (EMC)
effect~\cite{EuropeanMuon:1983wih}.
Thus, the properties of hadrons are also expected to be modified in the nuclear medium, 
that may be associated with the partial restoration of chiral symmetry.
Experimental evidence for this partial restoration of chiral symmetry has been
confirmed through the deeply bound pionic atoms~\cite{Suzuki:2002ae}, 
the low-energy pion-nucleus scattering~\cite{Friedman:2004jh}, and di-pion  
production in hadron-nucleus and photon-nucleus reactions~\cite{CHAOS:1996nql,CHAOS:2004rhl}. 
In the deeply bound pionic atom experiment~\cite{Suzuki:2002ae} and pion-nucleus 
scattering~\cite{Friedman:2004jh}, the analysis concluded that the pion decay constant 
(associated with the temporal part) is reduced in the nuclear medium.

From quantum chromodynamics (QCD) in the Standard Model (SM), we know that hadrons are composed 
of quarks and gluons. One can naturally expect that the quark and gluon dynamics  
is modified when hadrons are immersed in a nuclear medium. 
But, how are the internal structure of hadrons and the dynamics of the quarks 
and gluons modified in the nuclear medium? 
These questions, however, are still far remote to answer in terms of the first principle, QCD. 
Therefore, more studies are strongly required to understand the medium modifications
of hadron structure and properties. 
Motivated by this, many studies on the in-medium modifications of hadron properties and 
structure have been made both in theoretically~\cite{Hayano:2008vn} and 
experimentally~\cite{EuropeanMuon:1983wih,Suzuki:2002ae} (references therein). 
Until now, many theoretical studies have been made on the light and heavy mesons 
in the nuclear medium using various models and approaches, such as 
the QMC model~\cite{Saito:2005rv,Krein:2017usp,Zeminiani:2020aho,Cobos-Martinez:2022fmt}, 
the Dyson-Schwinger equation (DSE) based approach~\cite{Roberts:2000aa}, 
the holographic model~\cite{Kim:2022lng}, 
the QCD sum rules (QSR)~\cite{Park:2016xrw,Bozkir:2022lyk}, 
the Linear-sigma model (L$\sigma$M)~\cite{Suenaga:2019urn}, 
the Bethe-Salpeter equation-Nambu--Jona-Lasinio (BSE-NJL) 
model~\cite{Hutauruk:2018qku,Hutauruk:2021kej,Hutauruk:2019ipp,Hutauruk:2019was}, 
the instanton liquid model (ILM)~\cite{Nam:2008xx,Shuryak:1997vd}, 
and the hybrid light front-quark-meson coupling 
(LF-QMC) models~\cite{deMelo:2016uwj,deMelo:2014gea,deMelo:2018hfw}.

Among the models mentioned above, authors of 
Refs.~\cite{deMelo:2016uwj,deMelo:2014gea,deMelo:2018hfw} studied the 
weak-decay constants, DAs, and electromagnetic elastic form factors (EFFs) using the combined 
approach with the LFQM and the QMC models in the nuclear medium. 
However, their studies focused only on the $\pi$ and $\rho$ mesons with the light-light quark 
constituents. In this work, we extend the studies in a more systematic manner for the
light and heavy-light pseudoscalar as well as vector mesons. 
It is worth noting that, in the nuclear medium, the mesons  
with the heavy-light quark pair are subject to feel not only the scalar potential but 
also the vector potential in the QMC model. 
In this study, we employ a similar hybrid approach to that practiced 
in Refs.~\cite{deMelo:2016uwj,deMelo:2014gea,deMelo:2018hfw}, namely, 
the meson properties are computed in the LFQM, using the in-medium quark properties  
simulated in the QMC model. Note that, in the present study, 
we employ the Gaussian wave function in the LFQM, which 
is different from Refs.~\cite{deMelo:2016uwj,deMelo:2014gea,deMelo:2018hfw}, 
where they used the Bethe-Salpeter amplitude (BSA) with the mass regulator 
to tame the divergence in the loop integral by the quark propagators. 
Instead, here we introduce the quark potential 
in the Hamiltonian that describes the quark-antiquark interaction inside the meson.

The highlight of the present study is to study the dynamics of the light-light and 
heavy-light quark systems in the spin-0 pseudoscalar and spin-1 vector mesons, 
where the quarks in both mesons compose different symmetries and quantum numbers. 
The present study may also provide useful information on the properties of 
quarkonium in the nuclear medium~\cite{Krein:2017usp,Zeminiani:2020aho,Cobos-Martinez:2022fmt}. 
Such studies should be important for understanding the more complicated
quark dynamic systems like heavy baryons with bottom and/or charm quarks, as well as
exotic states in free space and in the nuclear medium.
These kinds of studies are relevant for the ongoing or planned experiments
in the modern, international facilities, such as PANDA and CBM collaborations at 
FAIR~\cite{PANDA:2009yku,Dbeyssi:2022zwz,Friman:2011zz,CBM:2016kpk,Prencipe:2015cgg}, 
PHENIX collaboration at RHIC~\cite{PHENIX:2005nhb}, BELLE and BELLE II collaboration at 
KEK~\cite{Prencipe:2018ugj}, LHCb collaboration at CERN~\cite{LHCb:2011zzp}, and 
J-PARC~\cite{Aoki:2021cqa}.

This article is organized as follows. 
In Sec.~\ref{sec:LFQM}, we first briefly introduce the effective Hamiltonian and 
light-front wave functions (LFWFs) used in the present LFQM approach. 
Then, we describe how the model parameters are fixed via the variational approach. Additionally,
we give the expressions for various properties of mesons, 
such as weak-decay constants, and DAs in free space in the LFQM formalism. 
In Sec.~\ref{sec:QMC}, we describe briefly the QMC model,   
so that we can input the medium effect in the LFQM calculation. 
In Sec.~\ref{sec:medium}, we present the formulas 
for the vector potentials, weak decay constants, and DAs for the light and 
heavy-light mesons in the nuclear medium. 
Section~\ref{sec:result} presents the results for the light and 
heavy-light meson properties in the nuclear medium. 
Section~\ref{sec:summary} is devoted to the summary and conclusion.

\section{Properties and structure of mesons in free space} \label{sec:LFQM}

In this section, we briefly describe the free space properties of the light and heavy-light mesons 
by employing the LFQM that is based on the constituent quark picture with light-front dynamics 
(LFD). Here, we first explain the key ideas of the LFQM, starting from the Hamiltonian and LFWF. 
We then show the expressions for the weak-decay constant and the DA in 
free space. We emphasize again that the present approach differs from the BSA approach of 
Refs.~\cite{deMelo:2016uwj,deMelo:2014gea,deMelo:2018hfw}, namely the present approach uses the 
vertex function that is regulated by the Gaussian functional form, 
and the meson states are built through the Bakamjian-Thomas (BT) construction~\cite{BT53,KP91}, 
which is a Poincar\'e invariant and guarantees that it is independent of any specific 
kinematics of any chosen frames~\cite{Arifi:2022qnd}.

\subsection{LFWF and effective Hamiltonian}

In the LFQM, a meson state is described as a bound state of the constituent quark and antiquark 
pair in the noninteracting representation following the BT construction, where the interaction is 
included in the meson mass operator to satisfy the Poincar\'e group 
structure~\cite{BT53,KP91}. 
Thus, the interaction is encoded in the mass eigenfunction. 
In the present approach, we apply a variational principle to deal with the mass eigenvalue problem, 
introducing a trial wave function in the Gaussian basis including the QCD-motivated effective 
Hamiltonian with linear confinement potential~\cite{CJ97}. 
The LFQM has been successfully applied to various studies of
free space meson properties in Refs.~\cite{ACJO22,CJ97,Choi:2009ai,Choi:2007yu,Choi:2017zxn} 
and references therein.

The meson state $\ket{{\cal M} (P,J, J_z)}$, as a bound state of the constituent 
quark $q$ and antiquark $\Bar{q}$ with meson momentum $P$ and total angular momentum $(J,J_z)$, 
can be written as
\begin{eqnarray}\label{eq:1}
\ket{{\cal M} (P,J, J_z)} 
&=& \int \left[ {\rm d}^3{\bf p}_q \right] \left[ {\rm d}^3{\bf p}_{\bar{q}} \right]  2(2\pi)^3 \delta^3 
\left({\bf P}-{\bf 
p}_q - {\bf p}_{\bar{q}} \right) 
\nonumber\\ && \times \mbox{} 
\sum_{\lambda_q,\lambda_{\bar{q}}} \Psi_{\lambda_q \lambda_{\bar{q}}}^{JJ_z}(x, {\bf k}_\perp)
\ket{q_{\lambda_q}(p_q) \bar{q}_{\lambda_{\bar{q}}}(p_{\bar{q}}) },
\quad\quad 
\end{eqnarray}
where ${\bf p_i}=(p_i^+,{\bf p}_{i \perp})$ and $\left[ {\rm d}^3{\bf p}_i \right] \equiv {\rm 
d}p_i^+ {\rm d}^2\mathbf{p}_{i \perp}/[2(2\pi)^3]$. 
Here, we define
$(p_q,\lambda_q)$ and $(p_{\bar{q}},\lambda_{\bar{q}})$
as the momentum and the helicity of quark $(i=q)$ and antiquark $(i={\bar{q}})$, respectively. The LF 
internal variables $(x, {\bf k}_\perp)$ are then denoted as $x=p^+_q/P^+$ and  
${\bf k}_{\perp} = {\bf p}_{q\perp} - x {\bf P}_\perp$.

The LFWF of the ground state meson in momentum space is defined by
\begin{eqnarray}\label{eq:5}
		\Psi^{JJ_z}_{\lambda_q\lambda_{\bar{q}} }(x, \mathbf{k}_{\bot}) = \Phi(x, \mathbf{k}_\bot)
		\  \mathcal{R}^{JJ_z}_{\lambda_q\lambda_{\bar{q}} }(x, \mathbf{k}_\bot),
\end{eqnarray}
where $\Phi(x, \mathbf{k}_\bot)$ and $\mathcal{R}^{JJ_z}_{\lambda_q\lambda_{\bar{q}}}(x, \mathbf{k}_\bot)$ 
are respectively 
the radial and spin-orbit wave functions, where the latter distinguishes the vector (V) and 
pseudoscalar (P) mesons. 
The $\mathcal{R}^{JJ_z}_{\lambda_q\lambda_{\bar{q}}}$ is obtained from the Melosh 
transformation~\cite{Melosh:1974cu} and 
has the covariant forms as
\begin{eqnarray}\label{eq:6}
	\mathcal{R}^{JJ_z}_{\lambda_q\lambda_{\bar{q}}} &=&  \frac{1}{\sqrt{2} \tilde{M}_0} 
	\bar{u}_{\lambda_q}^{}(p_q) \Gamma_{\rm M} v_{\lambda_{\bar{q}}}(p_{\bar{q}}),
\label{M0til}	
\end{eqnarray}
with ${\rm M}=$ P or V meson and $\tilde{M}_0 \equiv \sqrt{M_0^2 - (m_q -m_{\bar{q}})^2}$,
and the invariant meson mass $M_0^2$ is defined as
\begin{eqnarray}\label{eq:4}
	M_0^2 = \frac{\mathbf{k}_{\bot}^2 + m_q^2}{x}  + \frac{\mathbf{k}_{\bot}^2 + m_{\bar{q}}^2}{1-x}.
\end{eqnarray} 
In the above, the vertices for the pseudoscalar $\Gamma_{\rm P}$ and vector $\Gamma_{\rm V}$ are given by 
\begin{eqnarray}
\Gamma_{\rm P} &=& \gamma_5, \\
\Gamma_{\rm V} &=& -\slashed{\epsilon}(J_z) + \frac{\epsilon \cdot (p_q-p_{\bar{q}})}{ M_0 + 
m_q + m_{\bar{q}}},
\end{eqnarray}
where the polarization vectors $\epsilon^\mu(J_z)=(\epsilon^+, 
\epsilon^-,\bm{\epsilon}_{\perp})$ 
are defined by
\begin{eqnarray}\label{eq:7}
\epsilon^\mu(\pm 1) &=& \left( 0, \frac{2\bm{\epsilon}_\perp(\pm) \cdot {\bf P}_\perp}{P^+}, 
\bm{\epsilon}_\perp(\pm)\right),
\nonumber\\
\epsilon^\mu(0) &=& \frac{1}{M_0}\left(P^+, \frac{-M^2_0 + {\bf P}^2_\perp}{P^+}, {\bf 
P}_\perp\right),
\end{eqnarray}
with $\bm{\epsilon}_\perp(\pm 1) = \mp \frac{1}{\sqrt{2}} \left( 1, \pm i \right)$. We note that the 
the spin-orbit wave function is normalized to unity. 

In the meson ground state, the trial radial wave function in Gaussian basis is given by 
\begin{eqnarray}\label{eq:9}
	\Phi_{1S} (x, \mathbf{k}_\bot) &=& \frac{4\pi^{3/4}}{ \beta^{3/2}} 
	\sqrt{\frac{\partial k_z}{\partial x}} e^{-{\bf k}^2/ 2\beta^2},
\end{eqnarray}
with $\beta$ as the variational parameter that is related to the size of the wave
function, and the Jacobian factor in Eq.~(\ref{eq:9}) is expressed by
\begin{equation}
\frac{\partial k_z}{\partial x} = \frac{M_0}{4x(1-x)} \left[ 1 - \frac{ (m_q^2 - m_{\bar{q}}^2)^2}{M_0^4} 
\right],
\end{equation}
which takes account of the variable transformation, $(k_z,\mathbf{k}_\perp)$ to 
$(x,\mathbf{k}_\perp)$, 
where $k_z = \left( x - 1/2 \right) M_0 + (m^2_{\bar{q}} -m^2_q)/{2M_0}$. The LFWF is normalized by
\begin{eqnarray}\label{eq:10}
 \int \frac{{\rm d}x {\rm d}^2 \mathbf{k}_\bot}{2(2\pi)^3}  \abs{ \Psi (x, \mathbf{k}_\bot) }^2 =1.
\end{eqnarray}

Following the BT construction, the quark and antiquark interactions are included in the meson mass 
operator to 
compute the mass eigenvalue $M_{q\bar{q}}$, namely,
\begin{eqnarray}\label{LFeq:1}
(H_0 + V_{q\bar{q}} ) \ket{\Psi_{q\bar{q}}} = M_{q\bar{q}} \ket{\Psi_{q\bar{q}}},
\end{eqnarray}
where $\Psi_{q\bar{q}}$ is the eigenfunction for the meson.
In the calculation, we use the relativistic kinetic energy of the quark and antiquark
given by $H_0=\sqrt{m_q^2 + {\bf p}^2_q} + \sqrt{m_{\bar{q}}^2 + {\bf p}^2_{\bar{q}}}$, and
the quark-antiquark potential is defined by $V_{q\bar{q}}=V_{\rm Conf} + V_{\rm 
Coul} + 
V_{\rm Hyp}$, that consists of the confining, Coulomb-like, and hyperfine potentials, given by
\begin{eqnarray}
V_{\rm Conf} &=& a + br, \label{vconf}\\
V_{\rm Coul} &=& -\frac{4\alpha_s}{3r}, \label{vcoul}\\
V_{\rm Hyp} &=& \frac{32\pi \alpha_s \expval{\mathbf{S}_q \cdot \mathbf{S}_{\bar{q}}} }{9m_q m_{\bar{q}}} 
\delta^3(r), \label{vhype}
\end{eqnarray}
where $a$ and $b$ are parameters for the linear confining potential, and $\alpha_s$ is the strong 
running coupling, 
which is taken as a constant parameter in the free space.  
(Its in-medium modification will be explained later.)
The $\Braket{\mathbf{S}_q \cdot \mathbf{S}_{\bar{q}} }$ yields the values of $1/4$ and $-3/4$ 
for the vector and pseudoscalar mesons, respectively.

To obtain the mass and wave function of the meson ground state, we perform the variational 
analysis. 
The mass eigenvalue of the meson can be computed by
$M_{q\bar{q}}=\Braket{\Psi_{q\bar q}| H_{q\bar q}|\Psi_{q\bar q} }= \Braket{ \phi_{1S}| H_{q\bar 
q}|\phi_{1S} }$,
where the unity of the spin-orbit wave function has to be employed. We then have an analytic mass 
formula~\cite{Choi:2009ai}, 
\begin{eqnarray}
	M_{q\bar{q}} &=& \frac{\beta}{ \sqrt{\pi}} \sum_{i=q,{\bar{q}}} z_i e^{z_i/2}  
K_1\left(\frac{z_i}{2}\right) + a + \frac{2b}{\beta\sqrt{\pi}}  \nonumber\\
	& &  - \frac{8\alpha_s \beta}{3\sqrt{\pi}} +  \frac{32 \alpha_s \beta^3 \left<\mathbf{S}_q 
\cdot 
\mathbf{S}_{\bar{q}}\right>}{9 \sqrt{\pi} m_q m_{\bar{q}} },\quad 
\label{Mmass}
\end{eqnarray}
where $z_i = m_i^2/\beta^2$, $K_n$ the modified Bessel function of the second kind of order-$n$. 
The model parameters including the variational parameters $\beta$ are obtained by performing the 
variational principle and by imposing $\alpha_s$ to be the same for all mesons. 
The detailed procedures can be found in Ref.~\cite{CJ97}. 
Since our main purpose of the present work is to investigate the in-medium modifications 
of the meson properties and structure, we use the same parameters as those in free space, 
which are adapted from Ref.~\cite{CJ97}. 
The explicit values of the parameters used in the present LFQM will be presented in 
Sec.~\ref{sec:result}.

\subsection{Weak-decay constant and distribution amplitude}

Here, we review the properties and structure of the mesons in free space 
in the LFQM, such as the weak-decay constants and the DAs for the pseudoscalar and vector
mesons~\cite{Choi:2007yu}. 
The weak-decay constants are related to the normalization of 
the leading twist quark DAs of the corresponding mesons 
that dictate the probability of the valence-quark distributions, 
and they can be extracted through the hard exclusive reaction processes~\cite{Chernyak:1983ej, Lepage:1980fj}.  

The weak-decay constants of the pseudoscalar meson $f_{\rm P}$ 
and vector meson $f_{\rm V}$ are defined by
\begin{eqnarray}\label{eq:17}
	\bra{0} \bar{q} \gamma^\mu \gamma_5 q \ket{{\rm P}(P)} &=& i f_{\rm P} P^\mu, \nonumber\\
	\bra{0} \bar{q} \gamma^\mu q \ket{{\rm V}(P,J_z)} &=& f_{\rm V} M_{\rm V} \epsilon^\mu(J_z),
\end{eqnarray}
where $\epsilon^\mu (J_z)$ and $M_{\rm V}$ are the polarization vector and the mass of the vector 
meson, respectively. The explicit form of the decay constants in the LFQM is given by
\begin{eqnarray}\label{eq:18}
	f_{\rm M} &=& 2\sqrt{6}\int_0^1 {\rm d}x\int \frac{ {\rm d}^2 \mathbf{k}_\bot}{2(2\pi)^3}  
	\frac{ {\Phi}(x, \mathbf{k}_\bot) }{\sqrt{\mathcal{A}^2 + \mathbf{k}_\bot^2}} 
~\mathcal{O}_{\rm M}, 
\quad \quad 
\end{eqnarray}
where the operators $O_{\rm M}$ are defined by
\begin{eqnarray}
    \mathcal{O}_{\rm P} &=& \mathcal{A}, \label{eq:operator1}\\ 
    \mathcal{O}_{\rm V} &=& \mathcal{A}+ \frac{2 \mathbf{k}_\bot^2}{D_{0}}, \label{eq:operator2}
\end{eqnarray}
with $\mathcal{A} =  (1-x) m_q +  x m_{\bar{q}}$ and $D_{0}= M_0 + m_q + m_{\bar{q}}$.


The DAs of mesons are defined from the light-like separated gauge invariant free space-to-meson 
matrix elements, which can be interpreted as the probability amplitudes to find the hadron in a 
state with a minimum number of the Fock constituents and small transverse momentum separation~\cite{Choi:2007yu}.
The leading twist DAs for the pseudoscalar and vector mesons are obtained from the plus component 
of the currents, which are respectively given by
\begin{eqnarray}
A^+_P &=& \bra{0} \bar{q}(z) \gamma^+\gamma_5  q(-z) \ket{{\rm P}(P)}, \nonumber\\
&=& i f_{\rm P} P^+ \int^1_0 {\rm d}x\ {\rm e}^{i\zeta P \cdot z}  \phi_{\rm 
P}(x)\biggr|_{z^+=z_\perp=0},\\
A_V^+ &=& \bra{0}\bar{q}(z)\gamma^+ q(-z)\ket{{\rm V}(P,0)},\nonumber \\
&=& f_{\rm V} M_{\rm V} \epsilon^+(0) \int^1_0 {\rm d}x\  {\rm e}^{i\zeta P\cdot z}  \phi_{\rm 
V}(x)\biggr|
_{z^+=z_\perp=0},
\end{eqnarray}
where $\zeta=2x-1$. In the LFQM, the $\phi_{\rm M}(x)$ can be obtained by the transverse 
momentum integration of the LFWF: 
\begin{eqnarray}\label{LFeq:18}
	\phi_{\rm M}(x) &=& \frac{2\sqrt{6}}{ f_{\rm M}} \int \frac{ {\rm d}^2 
\mathbf{k}_\bot}{2(2\pi)^3}  
	\frac{ {\Phi}(x, \mathbf{k}_\bot) }{\sqrt{\mathcal{A}^2 + \mathbf{k}_\bot^2}} 
~\mathcal{O}_{\rm M}.\quad 
\end{eqnarray}
We note that the DAs are normalized as
\begin{equation} \label{eq:23}
	\int_{0}^{1} {\phi}_{\rm M}(x)\ {\rm d}x = 1.
\end{equation}

\section{Quark properties in nuclear matter} \label{sec:QMC}

Before proceeding to study the in-medium effect on the meson properties, we briefly explain the QMC 
model, where the parameters are fixed to reproduce the saturation properties of nuclear matter. 
The QMC model is a quark-based model of nuclear matter and nuclei with the relativistic mean field 
approximation~\cite{Guichon:1987jp,Guichon:1995ue,Saito:1996sf,Guichon:2018uew}. 
In this model, the Lorentz-scalar-isoscalar $\sigma$, Lorentz-vector-isoscalar $\omega$, and 
Lorentz-vector-isovector $\rho$ mean fields generated from the surrounding nuclear medium are 
directly coupled to the confined light $u$ and $d$ valence quarks. 
In this regard, the mean fields modify the light quark masses and 
energies, and thus modify the internal structure of nucleons in the nuclear medium. 
We note that the QMC model is different from the usual relativistic mean-field (RMF) 
approach, where the latter mean-field couples to the point-like 
nucleons~\cite{Serot:1997xg,Horowitz:2020evx}. 
Thus, in the QMC model, the coupling constants between the light quarks and mean fields 
are fixed to be the same for all the light quarks in any hadrons, once constrained by the 
nuclear matter saturation properties.

\subsection{Quark-based relativistic mean-field model: QMC model}

The effective Lagrangian density for the symmetric nuclear matter (SNM) at the hadronic level 
reads~\cite{Guichon:1987jp,Saito:2005rv,Guichon:1995ue,Saito:1996sf,Guichon:2018uew} 
\begin{eqnarray}
\mathscr{L}_{\mathrm{QMC}}  &=& \mathscr{L}_{\rm nucleon} + \mathscr{L}_{\rm meson} + 
\mathscr{L}_{\rm int},
\end{eqnarray}
and they are given by
\begin{eqnarray}
\mathscr{L}_{\rm nucleon} &=& 
 \bar{\psi}[i\slashed{\partial} - m_N ]\psi, 
\\
\mathscr{L}_{\rm meson} &=& \frac{1}{2} ( \partial_\mu \hat{\sigma}\partial^\mu\hat{\sigma} - 
m_\sigma^2\hat{\sigma}^2  ) \nonumber\\
& & - \frac{1}{2} \left[ \partial_\mu \hat{\omega}_\nu (\partial^\mu \hat{\omega}^\nu - 
\partial^\nu 
\hat{\omega}^\mu ) - m_\omega^2 \hat{\omega}^\mu \hat{\omega}_\mu\right],\quad\quad 
\end{eqnarray}
where $\psi$, $\hat{\sigma}$, and $\hat{\omega}$ represent the nucleon, $\sigma$, and $\omega$ 
field operators, respectively. In the present work, we do not include the $\rho$ meson field, 
since we consider the isospin-symmetric nuclear matter in which the contribution from the $\rho$ 
mean-field is vanishing in the Hartree mean-field approximation. The corresponding interaction Lagrangian density is given by
\begin{eqnarray}
\mathscr{L}_{\rm int} =\tilde{g}_\sigma^N(\hat{\sigma}) \bar{\psi} \psi \hat{\sigma}  - g_\omega^N 
\hat{\omega}^\mu 
\bar{\psi} \gamma_\mu \psi,
\end{eqnarray} 
with the $\sigma$-field dependent $N\sigma$ coupling constant $\tilde{g}_\sigma^N(\hat{\sigma})$ 
and the $N\omega$ coupling constant $g_\omega^N$. 
We may also write the Lagrangian density as
\begin{eqnarray}
\mathscr{L}  &=&  \bar{\psi}[i\slashed{\partial} - m_N^*(\hat{\sigma}) - g_\omega^N \hat{\omega}^\mu \gamma_\mu  ]\psi + \mathscr{L}_{\rm meson},
\end{eqnarray}
where the nucleon effective mass at a given density is defined by
\begin{eqnarray}
\label{eqmNmed}
m_N^*(\hat{\sigma}) = m_N - \tilde{g}_\sigma^N(\hat{\sigma}) \hat{\sigma}. 
\end{eqnarray} 
The $\tilde{g}_\sigma^N(\Hat{\sigma})$ coupling appears in the 
nucleon effective mass and it modifies the nucleon mass  
in a nonlinear manner in $\sigma$ field, 
while the $g^N_\omega$ coupling modifies the nucleon 
four-momentum. It is important to emphasize that in the QMC model, 
the couplings of the meson fields are applied directly to 
the light quarks and we solve the Dirac equation of the light quarks in the presence 
of the meson fields generated by the surrounded nuclear medium self-consistently. 
The quark-meson coupling constant with the $\sigma$ field  
is encoded in $\tilde{g}_\sigma^N(\hat{\sigma})$ at the hadron level.

In the mean-field approximation, we replace the meson field operators with their 
constant mean field expectation values as $\hat{\sigma} \to \sigma = \expval{\hat{\sigma}}$ and 
$ \hat{\omega}_\mu \to \delta_{\mu,0}\, \omega = \expval{\hat{\omega}_\mu}$, where we consider the 
nuclear matter at rest so that only the time component of $\omega_\mu$ survives.

At the nucleon level, the equations of motion of the meson fields can be obtained as
\begin{eqnarray} \label{eq:eom}
(\Box + m_\sigma^2)\sigma &=& \left(-\frac{\partial m_N^*(\sigma)}{\partial \sigma}\right) 
(\bar{\psi}\psi)  = 
\tilde{g}_\sigma^N(\sigma) \rho_s, \quad\quad  \\
(\Box + m_\omega^2)\omega &=& g_\omega^N (\bar{\psi}\gamma^0\psi) = g_\omega^N (\psi^\dagger\psi)  
=g_\omega^N \rho,
\label{eq:eom2}
\end{eqnarray}
where in nuclear matter, the d'Alembert operator $\Box \to 0$ is made, and $\rho_s$ and $\rho$ are 
the nucleon scalar and vector (baryon) densities, respectively. The Dirac equation for
the nucleon is given by
\begin{eqnarray}
(i\slashed{\partial} - g_\omega^N \omega \gamma^0 - m_N^{*}(\sigma) )\psi = 0.
\end{eqnarray}
Note that the effective nucleon mass enters in this equation and Eq.~(\ref{eq:eom}) via, 
\begin{eqnarray}
- \frac{\partial  m_N^*(\sigma) }{\partial \sigma} &=& \tilde{g}_\sigma^N(\sigma) = g_\sigma^N 
C_N(\sigma), 
\end{eqnarray}
where
\begin{eqnarray}
C_N(\sigma) = \frac{S_N(\sigma)}{S_N(\sigma=0)}. 
\end{eqnarray}
The $C_N(\sigma)$ is the so-called scalar polarizability that describes the nucleon response to the 
external scalar field~\cite{Saito:2005rv}.
If the nucleon is assumed as a point-like particle, $C_N(\sigma)=1$.
In the QMC model, the meson-nucleon coupling $(g_\sigma^N, g_\omega^N)$ are defined by the 
quark-meson coupling 
$(g_\sigma^q$,  $g_\omega^q)$ as
\begin{eqnarray}
g_\sigma^N &=& \tilde{g}_\sigma^N(\sigma=0) = 3g_\sigma^q S_N(\sigma=0),\\
g_\omega^N &=& 3g_\omega^q,
\label{gNs}
\end{eqnarray}
where $S_N(\sigma)$ is computed in the MIT bag model. In the above, factor three reflects the fact 
that the nucleon 
is made of three light valence quarks.

From Eqs.~(\ref{eq:eom}) and~(\ref{eq:eom2}), the vector and scalar meson fields are  respectively 
calculated as
\begin{eqnarray}
\omega &=& \frac{g_\omega^N \rho}{m_\omega^2},\label{eq:omega}\\
\sigma &=& \frac{g_\sigma^N \rho_s}{m_\sigma^2} C_N(\sigma), \label{eq:sigma}
\end{eqnarray}
where the nuclear density $\rho$ and scalar density $\rho_s$ are given by
\begin{eqnarray}
\rho &=&  \frac{4}{(2\pi)^3} \int d^3\mathbf{k}\ \Theta(k_F - k) = \frac{2k_F^3}{3\pi^2}, 
\label{eq:kf}\\
\rho_s &=& \frac{4}{(2\pi)^3} \int d^3\mathbf{k}\ \Theta(k_F - k) 
\frac{m_N^*(\sigma)}{\sqrt{m_N^{*2}(\sigma) + 
k^2}},
\end{eqnarray}
with $k=|\mathbf{k}|$ and $\Theta(k_F - k)$ being the step function that guarantees the integral is 
performed up to the nucleon Fermi momentum $k_F$ that can be written in terms of the nuclear 
density $\rho$. The factor of four represents spin and isospin degeneracy. As seen in 
Eq.~(\ref{eq:sigma}), we solve the self-consistent equation for the $\sigma$ mean-field to 
determine its value at each nuclear density. 

Once we obtain the $\sigma$ and $\omega$ mean fields, it is straightforward to compute 
the total energy per nucleon, which is given by 
\begin{eqnarray}
\frac{E_{\rm tot}}{A} &=& \frac{1}{\rho} \biggl[ \frac{4}{(2\pi)^3}\int d^3\mathbf{k}\ \Theta(k_F - 
k) 
\sqrt{m_N^{*2}(\sigma) +k^2 } \nonumber\\
 & & + \frac{1}{2} g_\sigma^N C_N(\sigma) \sigma \rho_s  + \frac{1}{2}g_\omega^N \omega \rho 
\biggr].
\end{eqnarray}
Later, we determine the model parameters by the fit to the extracted nuclear matter saturation 
properties at the saturation density $\rho_0=0.15$ fm$^{-3}$ ($k_F=1.305$ fm$^{-1}$),  
such as the negative of binding energy ($E_{\rm tot}/A - m_N$), $-15.7$ MeV.

\subsection{MIT bag model} 

As discussed earlier, the nucleon-meson couplings are defined based on the quark-meson couplings. 
In the standard 
QMC model~\cite{Guichon:1987jp,Guichon:1995ue,Saito:1996sf,Guichon:2018uew}, this is done by using 
the MIT Bag model for the nucleon (hadrons) and solving the Dirac equations for the quarks
and
antiquarks with the presence of the meson mean fields in the nuclear medium, 
\begin{eqnarray}
V_\sigma^q &=& g_\sigma^q \sigma,\\
V_\omega^{q} &=& g_\omega^q \omega,
\end{eqnarray}
where $\sigma$ and $\omega$ are the same constant mean fields as discussed in Eqs.~(\ref{eq:omega}) 
and (\ref{eq:sigma}).
Here we consider $q=u$ or $d$ and $Q=s,c,$ or $b$ quark confined inside the bag of a hadron in 
symmetric nuclear matter at the position $z=(t,\mathbf{r})$ with $r=|\mathbf{r}| \leq$ bag radius.

The Dirac equations for the quark and antiquark in the presence of the mean-field potentials 
are given by
\begin{eqnarray}
\biggl[i\slashed{\partial} - (m_q - V_\sigma^q) \mp \gamma^0 V_\omega^q \biggr] 
\begin{pmatrix}
\psi_q (z)\\
\psi_{\bar{q}} (z)\\
\end{pmatrix} = 0, \\
\left[i\slashed{\partial} - m_Q\right]
\begin{pmatrix}
\psi_Q(z)\\
\psi_{\bar{Q}} (z)\\
\end{pmatrix}= 0,
\end{eqnarray}
where we assume the SU(2) symmetry for the light quarks, $m_q=m_u=m_d$. While the vector potential 
shifts the quark 
energy in the nuclear medium, the scalar potential modifies the quark mass as
\begin{eqnarray}
    m_q^* = m_q - V_\sigma^q,
\end{eqnarray}
so that the quark effective mass is reduced by the scalar $\sigma$ potential 
in the nuclear medium, and thus $m_q^*$ can be negative 
when a small free space mass value $m_q$ is used, 
but it is nothing but the reflection of the attractive 
potential, and thus a naive interpretation of ``mass'' should not be applied. 
It should be noted again that the mean fields are coupled only to the light quarks and antiquarks.
In the QMC model, the masses of the quark $Q=s,c,b$ are assumed to be the same in the nuclear 
medium as in free space ($m_Q^* = m_Q$) [but energies may be modified by the modifications of 
the hadron bag radius when a light quark is contained in the hadron], 
since the $\sigma$ mean-field is not coupled to the 
heavier quarks in the model, based on the fact that the heavy quark chiral condensates 
are expected to be modified only slightly in the cold nuclear medium, or, their 
masses are mostly due to the Higgs mechanism.
(See the introduction part of Ref.~\cite{Tsushima:2020gun} for detailed discussions   
and some evidence, as well as references on this issue.)

Here, we obtain the static solution for the ground state quark or antiquark, where the Hamiltonian 
is time-independent, and the wave function can be written as
\begin{eqnarray}
\psi(z) = \psi(r) \exp{-i\varepsilon^* t/R^*},
\end{eqnarray}
with the in-medium bag radius $R^*$. The eigenenergies in units of $1/R^*$ are given by
\begin{eqnarray}
\begin{pmatrix}
 \varepsilon^*_q \\
\varepsilon^*_{\bar{q}} \\
\end{pmatrix} &=& \Omega_q^* \pm R^* V_\omega^q,
\end{eqnarray}
with 
\begin{eqnarray}
    \Omega^*_q &=& \sqrt{x_q^{*2} + (m_q^* R^*)^2},
\end{eqnarray}  
where $x_q^*$ is the lowest mode bag eigenvalue. The normalized ground state quark eigenfunction can 
be written as
\begin{eqnarray}
\psi(z) = \frac{N {\rm e}^{-i\varepsilon^* t/R^*}}{\sqrt{4\pi}} \begin{pmatrix}
    j_{0}(x_q^* r/R^*) \\ i\beta_q^* j_{1}(x_q^* r/R^*)\ {\bm \sigma} \cdot \hat{r} 
\end{pmatrix}  \chi_m,
\end{eqnarray}
where $\chi_m$ is the spin function, the spherical Basel functions are given by
\begin{eqnarray}
j_0(r) &=& \frac{\sin r}{r},\\
j_1(r) &=& \frac{\sin r}{r^2} -  \frac{\cos r}{r}.
\end{eqnarray}
The normalization $N$ is obtained from
\begin{eqnarray}
\int_0^{R^*} {\rm d}^3\mathbf{r}\  \psi^\dagger(r) \psi(r) = 1,
\end{eqnarray}
where the integration is up to the radius $R^*$ meaning that the quarks only exist (confined) in 
the 
spherical 
cavity. We then obtain 
\begin{eqnarray}
N^{-2} = 2 R^{*3} j_0^2(x_q^*) \frac{[\Omega_q^*(\Omega_q^*-1) + m_q^* R^*/2]}{x_q^{*2}}.
\end{eqnarray}
The eigenfrequency of the quark is obtained by imposing the continuity of the quark 
eigenfunction at the bag boundary $r=R^*$ which yields the relation, 
\begin{eqnarray}
j_0(x_q^*) = \beta_q^* j_1(x_q^*),
\end{eqnarray}
with 
\begin{eqnarray}
\beta_q^* = \sqrt{\frac{\Omega_q^* - m_q^* R^*}{\Omega_q^* + m_q^* R^*}}.
\end{eqnarray}
By solving the above equation, we obtain the lowest positive eigenvalue of $x_q^*$. The effective 
nucleon mass in the MIT bag model is computed as
\begin{eqnarray}
m_N^*(\sigma) = \frac{3\Omega^*_q - Z_N}{R^*} +\frac{4\pi R^{*3}}{3} B,
\label{mNsmass}
\end{eqnarray}
where $Z_N$ is the corrections from the gluon fluctuation and center-of-mass 
motion~\cite{Guichon:1995ue} and $B$ 
is the density-independent bag pressure. 
It is worth noting that the bag volume energy provides an inward pressure while the kinetic energy 
gives an outward 
pressure. 
As a consequence, there is an equilibrium where the nucleon mass is minimized and stabilized by  
\begin{eqnarray}
\left.\frac{{\rm d} m_N^*(R^*)}{{\rm d} R^*}\right|_{R^*=R_N^*} = 0,
\end{eqnarray} which determines the $R^*_N$ and mass of the nucleon, self-consistently 
with $x_q^*$. 
We also clarify 
that the bag radius ($R^*_N$) is not observable, which is different from the nucleon radius. 
So, to estimate the nucleon radius (in the nuclear medium as well as in free space),
one needs to compute it from the quark wave function. 
Lastly, the quantity associated with the scalar polarizability $C_N(\sigma)$,  
$S_N(\sigma) = S_N(\sigma=0)\, C_N(\sigma)$ in the $\sigma$-dependent $N\sigma$ coupling constant 
$\tilde{g}_\sigma^N(\sigma)$, is computed as
\begin{eqnarray}
S_N(\sigma) &=& \int_0^{R^*}  {\rm d}^3\mathbf{r}\ \bar{\psi}(r)\psi(r), \nonumber\\
&=& \frac{\Omega_q^*/2 + m_q^* R^*(\Omega_q^*-1)}{\Omega_q^*(\Omega_q^*-1) + m_q^* R^* /2}.
\end{eqnarray}
This provides one of the origins for the novel saturation properties of nuclear matter starting 
from the quark degrees of freedom, basing on the quark structure of the nucleon. 
With this ingredient, we do not have to introduce the nonlinear couplings of the meson fields in 
the effective Lagrangian density at the hadronic level to obtain a reasonable
incompressibility value in the range $K\simeq$ 200 - 300 MeV, which is usually adopted in
many sophisticated relativistic mean field models at the hadron level~\cite{Serot:1997xg}
for obtaining a reasonable value $K$.

\section{Properties and structure of meson in nuclear medium} \label{sec:medium}

In this section, we explore the mechanism of how the in-medium modified quark properties 
simulated in the QMC model affect the meson properties in nuclear medium,
such as weak-decay constants and DAs described in the LFQM. 
Two important inputs obtained from the QMC model applied are: (i) the in-medium light quark 
effective mass modified by the scalar $\sigma$ mean-field, and (ii) the in-medium light quark 
energy modified by the vector potential.
Since we work on the light and heavy-light
mesons, the medium effects vary for each type of meson depending on the quark
constituents of mesons.
We will explain this in the following. 
In this work, the Gaussian parameter $\beta$ is not assumed to be modified in the medium.
This may be justified by the fact that it is associated with a short-range scale in the meson wave function~\cite{deMelo:2014gea}. 
One may also expect that $\beta$ decreases in the medium as the meson radius increases. 
In that case, the decay constant will decrease much faster in the medium as studied by one of us in the previous work~\cite{deMelo:2018hfw}.\footnote{ The parameter $\beta$ is related to the regulator mass $m_R$ in Ref.~\cite{deMelo:2018hfw}.}
A quick decrease in the decay constant is due to the fact that it is more sensitive to the origin of the wave function. 
However, such a density dependence is not well understood either and will add only model ambiguities. Therefore, we assume the density-independent $\beta$ in this exploratory study.

The light quark or light antiquark mass is modified by the scalar potential as
\begin{eqnarray}
m_q^* = m_q - V_\sigma^q,
\end{eqnarray}
while the light quark and light antiquark energies $p^{*0}_{i}$ are modified by the
vector potential:
\begin{eqnarray}
p^{*0}_{i} =
\begin{cases}
E_q^* + V_\omega^q, & {\rm for\ light\ quark,}\\
E_{\bar{q}}^* - V_\omega^{q}, & {\rm for\ light\ antiquark,}\\
\end{cases}
\end{eqnarray}
with $i=q,\bar{q}$ and $E^*_q = E^*_{\bar{q}} = \sqrt{m_q^{*2} + \mathbf{p}^2_q}$.
Then, the total energy of meson $P^{*0}$ is given by
\begin{eqnarray}
P^{*0} = 
\begin{cases}
E_M^*, &{\rm for\ } (q\bar{q}),\\
E_M^* + V_\omega^q, &{\rm for\ } (q\bar{Q}),\\
E_M^* - V_\omega^{q}, &{\rm for\ } (Q\bar{q}),\\
\end{cases}
\end{eqnarray}
where $E_M^* = \sqrt{M^{*2} + \mathbf{P}^2}$ with $M^*$ being the in-medium meson
mass, and the vector potential cancels in the case of $(q\Bar{q})$ mesons,
and no vector potentials are acted for $Q$. The variable $x$, defined by the
plus-component ratio of the quark to meson momenta in free space,
\begin{eqnarray}
x = \frac{p^+_q}{P^+}
= \frac{p^0_q + p^3_q}{P^0 + P^3} = \frac{E_q + p^3_q}{E_M + P^3},
\end{eqnarray}
will be modified in the medium by the scalar and vector potentials.

For the $q\bar{q}$ mesons, we define
\begin{eqnarray}
p_q^{*+} = p_{\bar{q}}^{*+} &\equiv& E_q^* + p^{*3}_q,\\
P^{*+} &\equiv& E^*_M + P^{*3},\\
x^* &\equiv& \frac{p^{*+}_q}{P^{*+}},
\end{eqnarray}
and the longitudinal light quark ($q$) momentum is modified by
\begin{eqnarray}
x &\to&
 \tilde{x}^*
= \frac{p_q^{*+} +  V_\omega^q }{P^{*+}} = x^* + \frac{V_\omega^q }{P^{*+}},
\end{eqnarray}
where the vector potentials for the $q$ and $\bar{q}$ cancel for the $q\bar{q}$ mesons.
With this new definition, we have to shift the longitudinal quark momentum newly denoted
by $x \to \Tilde{x}^*$ because $(p_q^+,P^+) \ne (p_q^{*+}, P^{*+})$ in computing the weak
decay constant in the medium,
\begin{eqnarray}\label{eq:long}
\tilde{x}^* = x^* + \frac{V_\omega^q }{P^{*+}},
\end{eqnarray}
and then, the weak-decay constant is calculated as
\begin{eqnarray}\label{eq:18a}
	f_{\rm M}^* &=& 2\sqrt{6} \int_{-\frac{V_\omega^q }{P^{*+}}}^{1-\frac{V_\omega^q 
}{P^{*+}}} {\rm d}x^* 
\int \frac{ {\rm d}^2 \mathbf{k}_\bot}{2(2\pi)^3}  \nonumber\\
 & & \times \frac{ {\Phi}(\tilde{x}^*, \mathbf{k}_\bot) }{\sqrt{\mathcal{A}(\Tilde{x}^*)^{2} + 
\mathbf{k}_\bot^2}} 
~\mathcal{O}_{\rm M}(\tilde{x}^*,\mathbf{k}_\perp). \quad 
\end{eqnarray}
As seen in Eq.~(\ref{eq:18a}), the integration limits of $x^*$ are shifted by 
${V_\omega^q}/{P^{*+}}$. We then 
evaluate the equation by integration over $x^*$, which will give the same result as the decay 
constant expression in Eq.~(\ref{eq:18}). This can be understood because the upper limit of 
the integration will cancel with the lower limit of integration when integrating over $x^*$.
So, the final expression for the decay constant in the medium, using the variable $\Tilde{x}^*$, is 
given by\footnote{ $\Tilde{x}^*$ is a dummy variable
to emphasize the difference with $x$.}
\begin{eqnarray}\label{eq:19a}
	f_{\rm M}^* &=& 2\sqrt{6} \int_{0}^{1} {\rm d}\tilde{x}^* \int \frac{ {\rm d}^2 
\mathbf{k}_\bot}
{2(2\pi)^3}  \nonumber\\
 & & \times \frac{ {\Phi}(\tilde{x}^*, \mathbf{k}_\bot) }{\sqrt{\mathcal{A}(\Tilde{x}^*)^{2} + 
\mathbf{k}_\bot^2}} 
~ \mathcal{O}_{\rm M}(\tilde{x}^*,\mathbf{k}_\perp). \quad 
\end{eqnarray}
This equation shows that the scalar potential modifies the weak decay constant in the medium 
through the effective 
quark mass $m_q^*$~\cite{deMelo:2016uwj}, while the vector potential modifies the energies but they cancel each other between the quark and antiquark as in
Eq.~(\ref{eq:19a}). Note that, since $f_{\rm M}^*$
is computed with the plus current $J^+ =J^0 +J^3$ in the LFQM, we cannot separate the time and 
space components, as discussed in Refs.~\cite{Kirchbach:1997rk,Nam:2008xx}.

For the $q\bar{Q}$ and $Q\bar{q}$ mesons,
both the scalar and vector potentials contribute explicitly to the meson
four-momenta and weak-decay constants. The longitudinal momenta for
the $q$ and $\bar{q}$ in medium are given by
\begin{eqnarray}
x &\to& 
\begin{cases}
 \tilde{x}^* = \dfrac{p_q^{*+} + V_\omega^q }{P^{*+} + V_\omega^q} =  \dfrac{x^* +
V_\omega^q/P^{*+}}{(1 + 
V_\omega^q/P^{*+})}, & {\rm for\ } (q\Bar{Q}),\\ 
 \\
  \tilde{x}^* = \dfrac{p_q^{*+} - V^q_\omega }{P^{*+} - V^q_\omega} =

\dfrac{x^*  - V^q_\omega/P^{*+}
}{(1 - V_\omega^q/P^{*+}) },  
  & {\rm for\ } (Q\Bar{q}). \nonumber
\end{cases}\\
\end{eqnarray}

In such cases, we can define the variable of the integration by
\begin{eqnarray}
{\rm d}x^* = (1 \pm V_\omega^q/P^{*+})\ {\rm d}\tilde{x}^*,
\end{eqnarray}
and the meson weak decay constants for the $q\bar{Q}$ and $Q\bar{q}$ mesons in the medium 
can be written as
\begin{eqnarray}\label{eq:vector}
 f_{\rm M}^* &=& 2\sqrt{6} \int_{0}^{1} {\rm d}\tilde{x}^* \int \frac{ {\rm d}^2 
\mathbf{k}_\bot}
{2(2\pi)^3}  \left(1 \pm \frac{V_\omega^q}{P^{*+}}\right) \nonumber\\
 & & \times \frac{ {\Phi}(\tilde{x}^*, \mathbf{k}_\bot) }{\sqrt{\mathcal{A}(\Tilde{x}^*)^{2} + 
\mathbf{k}_\bot^2}} 
~ \mathcal{O}_{\rm M}(\tilde{x}^*,\mathbf{k}_\perp). \quad 
\end{eqnarray}
Our final expressions for the weak-decay constants for the heavy-light mesons with vector potential 
in the medium are similar to those of the quark distribution in the medium with vector 
potential obtained in Ref.~\cite{Steffens:2004yb}.\footnote{The quark distribution with and without 
vector potential differs by the factor of $(\tilde{P}^{*+}/P^{*+})$~\cite{Steffens:2004yb}, 
which is equivalent to the $(1\pm V_q^\omega/P^{*+} \to 1)$ in this work.}
In the meson rest frame, we have 
$P^{*+} = E^*_M + P^{*3} = \sqrt{M^{*2} + \mathbf{P}^2} + P^{*3}= M^*$.
In our LFQM, the meson mass should be replaced by the 
interaction-independent invariant mass $M_0^*$. However, in the case of the weak decay constant for 
the $q\bar{Q}$ and $Q\bar{q}$ mesons, 
we obtain that the decay constants are frame-dependent because of 
the $V^q_\omega / P^{*+}$ appearance in 
Eq.~(\ref{eq:vector}). 
In fact, since $V^q_\omega$ is the time component of the vector mean field obtained in 
the nuclear matter rest frame, this should also be Lorentz transformed for the general frame with 
nonzero relative velocity case between the frames.

\begin{table}[b]
	\begin{ruledtabular}
		\renewcommand{\arraystretch}{1.2}
		\caption{The quark masses and $\beta$ parameters used in the present work in  
units of GeV. The LFQM parameters are given by $a=-0.724$ GeV, $b=0.18$ GeV$^{2}$, and 
$\alpha_s=0.313$ (this will be modified in medium, to be discussed later), which are adapted from Ref.~\cite{CJ97}.}
		\label{parameter}
		\begin{tabular}{cccccccc}
		 $m_q$ & $m_s$ & $m_c$ & $m_b$  &	$\beta_{q\bar{q}}$ & $\beta_{q\bar{s}}$ & 
		 $\beta_{q\bar{c}}$ & 
$\beta_{q\bar{b}}$  
\\ \hline
		 0.22 & 0.45 & 1.8 & 5.2 & 0.3659 & 0.3886 & 0.4679 & 0.5266 \\
		\end{tabular}
		\renewcommand{\arraystretch}{1}
	\end{ruledtabular}
\end{table}
\begin{table}[t]
	\begin{ruledtabular}
		\renewcommand{\arraystretch}{1.1}
		\caption{The predicted masses and weak-decay constants of the light and heavy-light pseudoscalar and vector mesons in comparison with experimental data, adapted from Ref.~\cite{CJ97}. The underlined experimental data represents the model's input. }
		\label{mass}
		\begin{tabular}{ccccc}
		    & $M_{\rm expt}$ [MeV] & $M_{\rm theo}$ [MeV] & $f_{\rm expt}$ [MeV] & $f_{\rm 
theo}$ [MeV] \\ 
\hline  
		 $\pi $ &    \underline{135} & 135  & 130 & 130 \\
		 $\rho$ &    \underline{770} & 770  & 216 & 247 \\
		 $K $   &    498 & 478  & 156 & 162 \\
		 $K^* $ &    892 & 850  & 217 & 256 \\
		 $D $   &   1865 & 1836 & 206 & 197 \\
		 $D^* $ &   2007 & 1998 & \dots & 239  \\
		 $B $   &   5279 & 5235 & 188 & 171 \\
		 $B^* $ &   5325 & 5315 & \dots & 186  \\
		\end{tabular}
		\renewcommand{\arraystretch}{1}
	\end{ruledtabular}
\end{table}

In Section~\ref{sec:result}, we will present the results by dropping the factor of $(1 \pm 
V_\omega^q/P^{*+})$ in the weak decay constants.
This can be understood as the average of the decay constant for the meson multiplet   
as $f^* = (f_{q\Bar{Q}}^* + f_{Q\Bar{q}}^* )/2.$
Analogously, this ``average'' will also be practiced for the calculation of the in-medium 
modifications of DAs.

\section{Numerical result}\label{sec:result}

In this section, we present our numerical results for the in-medium modifications of the weak-decay 
constants and DAs for the light and heavy-light pseudoscalar and vector mesons in symmetric nuclear 
matter. The results are shown in Figs.~\ref{fig:constant}-\ref{fig:da_diff}. 
Before discussing the obtained results, we first explain the model parameters 
of the LFQM and the QMC model.

\subsection{Meson mass and weak-decay constant\\ 
 in free space} 

The model parameters of the LFQM (in free space) are determined by fitting the 
mass spectra of the $\pi$ and $\rho$ mesons, as practiced in Ref.~\cite{CJ97}. 
The parameters determined are shown in Table~\ref{parameter}. 
With these parameters, the predicted masses and weak-decay constants for the light  
and heavy-light pseudoscalar and vector mesons are in reasonable agreement 
with the data, as shown in Table~\ref{mass}.
We note that a further improvement of the model can be done by modifying the trial wave function and effective Hamiltonian~\cite{ACJO22} and reference therein.
Using the same parameters given in Table~\ref{parameter}, 
we study the meson properties in symmetric nuclear matter 
with the help of the QMC model inputs for the in-medium light quark properties.

\subsection{Nuclear matter} 

In this section, we explain the QMC model parameters as well as those  
of the MIT bag model. The quark-meson coupling constants are determined by 
the fit to the equation of state (EoS) extracted by the empirical data.

\subsubsection{Bag parameter}
\begin{figure}[t]
	\centering
	\includegraphics[width=0.9\columnwidth]{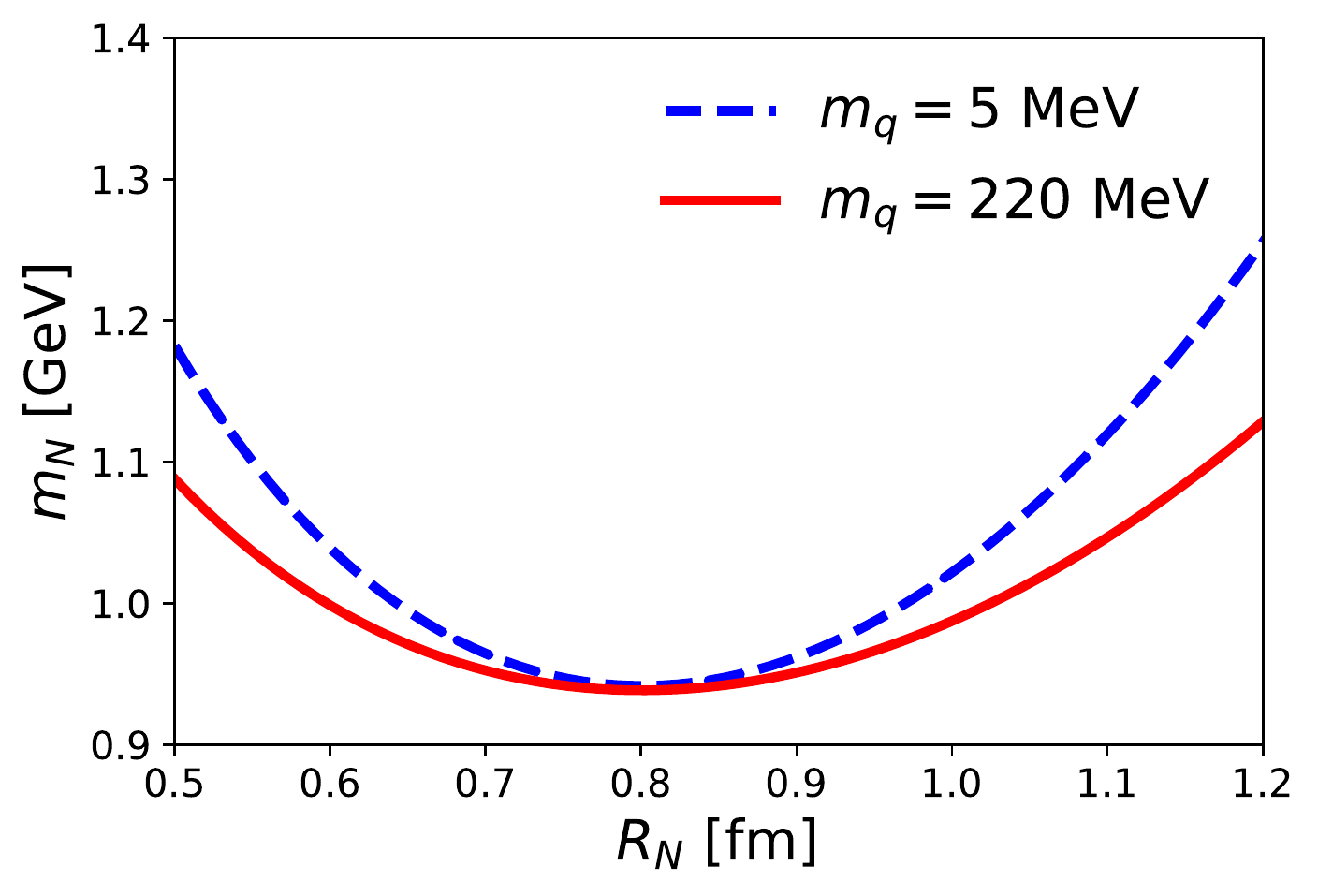}
	\caption{\label{fig:bag} 
Dependence of the nucleon mass $m_N$ on the nucleon bag radius $R_N$ for different values of the free space quark masses, 
 $m_q=$ 5 and 220 MeV. The physical nucleon mass $m_N = 939$ MeV (input) is 
achieved by the minimization condition, $\frac{{\rm d}m_N}{{\rm d}R}\bigr|_{R=R_N}=0$ with the 
input $R_N = 0.8$ fm, by varying unknown values to determine $B^{1/4}, Z_N$ and $x_q$. 
Once $B^{1/4}$ and $Z_N$ are determined by the minimization condition, $R_N$ dependence of $m_N$ is 
calculated with the fixed values of $B^{1/4}$ and $Z_N$, and with $x_q$ obtained by
solving $j_0 (x_q) = \beta_q j_1(xq)$ ($R_N$ dependent), and all the relevant values into
the MIT bag mass formula Eq.~(\ref{mNsmass}) but for the free space ($\sigma = 0$).
}
\end{figure}

\begin{figure*}[t]
	\centering
    \includegraphics[width=1.8\columnwidth]{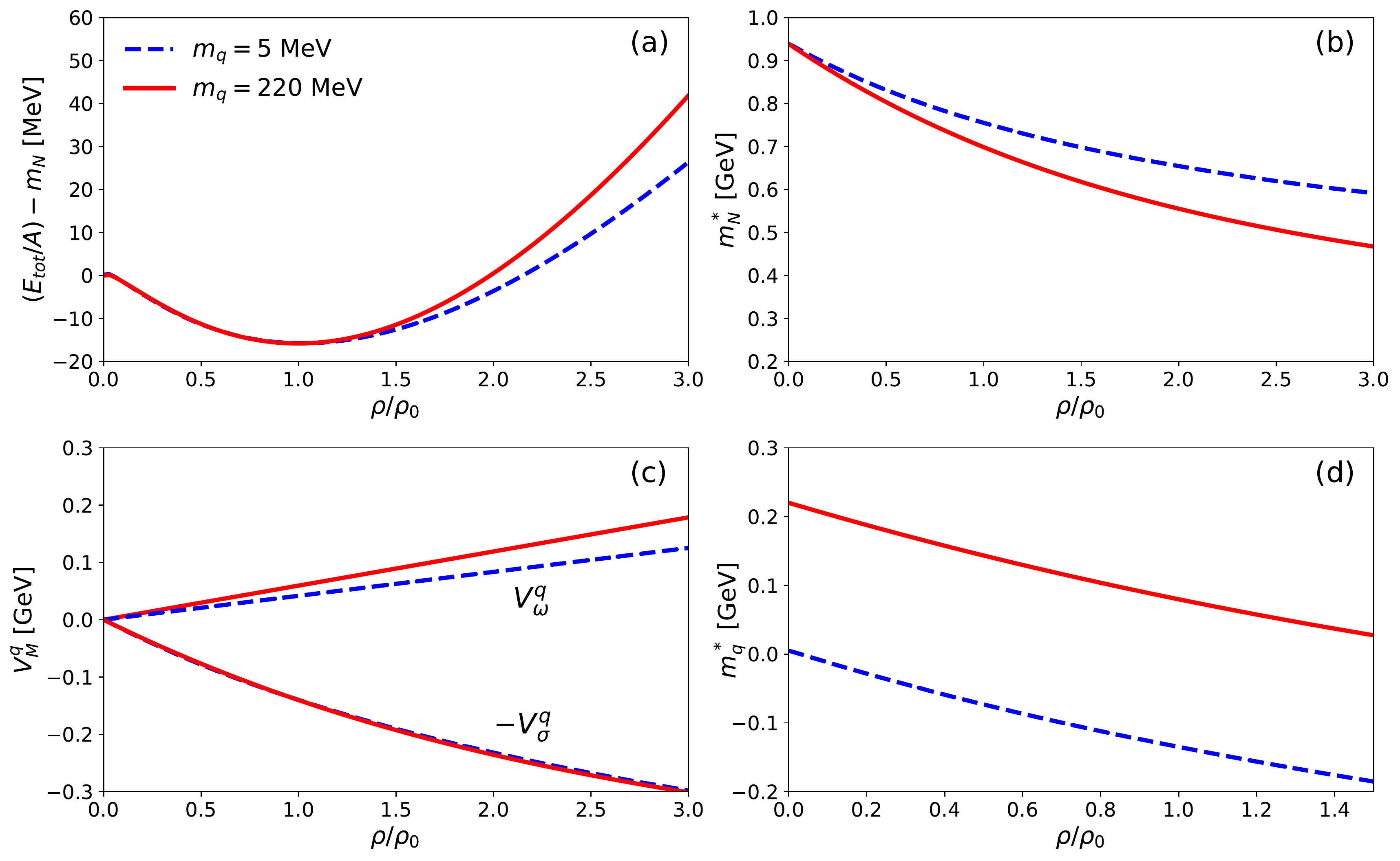}
	\caption{\label{binding_qmc} 
(a) The total energy per nucleon [$E_{\rm tot}/A - m_N$], minimized at the saturation
density $\rho_0 = 0.15$ fm$^{-3}$ with the negative of binding energy $-15.7$ MeV,
(b) the effective nucleon mass $m_N^*$, 
(c) the mean-field potentials of $V_\omega^q$ and $-V_\sigma^q$,
and (d) the light-quark effective mass with $m_q=$ 5 and 220 MeV.}
\end{figure*}

\begin{table}[b]
	\begin{ruledtabular}
		\renewcommand{\arraystretch}{1.2}
		\caption{The bag parameters are determined by reproducing the nucleon mass ($m_N = 
939$\ MeV) and radius ($R_N= 0.8$\ fm) in free space for two different values of the quark masses. The lowest positive eigenvalue $x_q$ and a constant $S_N(\sigma=0)$, associated with the scalar polarizability and used to define the $g^N_\sigma$ coupling constant, are also quoted. 
(See also Eq.~(\ref{gNs}).) }
		\label{tab:bag}
		\begin{tabular}{ccccc}
		$m_q$ [MeV]  &	$B^{1/4}$ [MeV] & $Z_N$ & $x_q$ & $S_N(\sigma=0)$\\ \hline
		5 & 170 & 3.295 & 2.052   & 0.483\\
		220 & 148 & 4.327 & 2.368 & 0.609 \\
		\end{tabular}
		\renewcommand{\arraystretch}{1}
	\end{ruledtabular}
\end{table}

In the QMC model, the bag model parameters $(Z_N$ and $B)$ are fitted to the nucleon mass and 
the bag radius in free space,
\begin{eqnarray}
m_N &=& 939\ {\rm MeV},\\
R_N &=& 0.8\ {\rm fm},
\end{eqnarray}
together with the mass stability/minimization condition which yield 
the bag pressure $B$ and $Z_N$ values in free space.
Here, we show the results for two different values of the quark masses with $m_q=$ 5 MeV and 
220 MeV. The values of the obtained bag parameters are summarized in Table~\ref{tab:bag}. 
The results for the $R_N$ dependence of nucleon mass in free space for two 
different quark masses are also shown in Fig.~\ref{fig:bag}. 
It clearly shows that the nucleon mass is minimized at a radius of 0.8 fm, as it should be 
by the stability/minimization for the both quark mass values.
It is worth noting that the $m_q=$ 5 MeV is the 
standard input value in the QMC model~\cite{Saito:2005rv}. 
In addition, when we use the constituent quark mass, the values of the bag parameters are 
insignificantly changing.

\subsubsection{Equation of state}

Here, we show the EoS of nuclear matter for $m_q=5$ and 220 MeV, where the bag parameters are 
determined in free space. As shown in Table~\ref{tab:coupling}, the obtained coupling 
constants $g_\sigma^N$ and $g_\omega^N$ for $m_q=220$ MeV are slightly larger 
than those for $m_q=5$ MeV, where both cases are determined by fitting 
the negative of binding energy $-15.7$ MeV at saturation density $\rho_0=0.15$ fm$^{-3}$ 
($k_F=1.305$ fm$^{-1}$).

\begin{table}[b]
	\begin{ruledtabular}
		\renewcommand{\arraystretch}{1.2}
		\caption{Coupling constants and the incompressibility $K$ calculated by the QMC model.
The bag model parameters for each quark mass value are determined in free space. 
The coupling constants are determined by the negative of binding energy $-15.7$ MeV 
at the saturation density $\rho_0=$ 0.15 fm$^{-3}$ ($k_F=$ 1.305 fm$^{-1}$). 
(See also Eq.~(\ref{gNs}).)
}
		\label{tab:coupling}
		\begin{tabular}{ccccc}
		$m_q$ [MeV] &	$(g_\sigma^N)^2/4\pi$ & $(g_\omega^N)^2/4\pi$ & $m_N^*$ [MeV] &  $K$ 
[MeV]\\ \hline
		5 & 5.39 & 5.30 & 755 & 279 \\
		220 & 6.40 & 7.57 & 699 & 321 \\
		\end{tabular}
		\renewcommand{\arraystretch}{1}
	\end{ruledtabular}
\end{table}

Figures~\ref{binding_qmc} (a) and (b) display the total energy per nucleon and effective nucleon 
mass versus nuclear density ratio $\rho/\rho_0$ for two different quark mass values, respectively. 
For the larger quark mass value, the larger incompressibility $K$ is obtained. 
In comparison with empirical values of $K$, we find that the obtained values of $K$ for 
both cases are consistent with the empirical range $K = 200-300$ MeV~\cite{Stone:2014wza}, 
although the $K$ for $m_q = 220$ MeV is slightly larger about 7\%. 
The density dependence of the energy per nucleon [$(E_{\rm tot}/A)-m_N$] for 
both quark mass cases have very similar values up to $\rho \simeq \rho_0$, 
and then they begin to show the difference as nuclear density increases 
in the range $\rho \gtrsim \rho_0$. 
Figure~\ref{binding_qmc} (b) shows that the effective nucleon mass for $m_q = 220$ MeV  
decreases faster as nuclear density increases than that for 5 MeV.
The values of incompressibility and the effective 
nucleon mass at saturation density $\rho_0$ for two $m_q$ values are given in 
Table~\ref{tab:coupling}.

\begin{figure}[t]
	\centering
    \includegraphics[width=0.9\columnwidth]{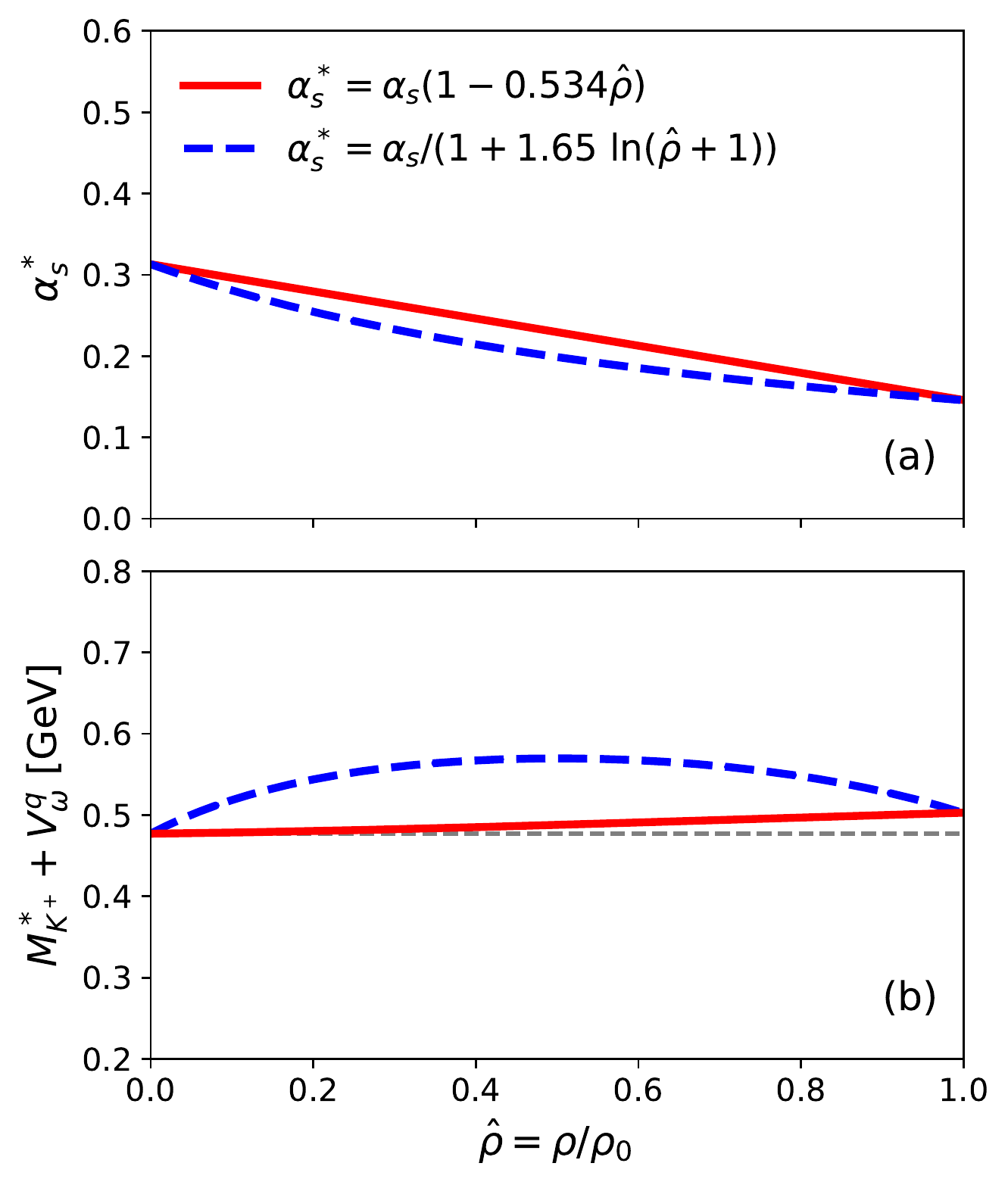}
	\caption{\label{fig:as_med} 
(a) Modifications of $\alpha_s^*$ in the nuclear medium with two different forms,
and (b) $K^+$ excitation energy $K^+$ at rest,
$\omega_{K^+} \equiv M^*_{K^+} + V^q_\omega$ ($M^*_{K^\pm} = M^*_{K,\bar{K}}$)
in the nuclear medium. We have determined the parameters to reproduce the
repulsive $K^+$ total potential, $\omega_{K^+} = M^*_{K^+} + V^q_\omega
= M_{K^+} + (20-30)$ MeV at $\rho_0$.
}
\end{figure}

Besides the energy per nucleon and effective nucleon mass, we also show the mean-field
potentials for the light quarks, which should also apply to the light quarks in the
mesons. The mean-field potentials, $V_\omega^q$ and $-V_\sigma^q$ are depicted in
Fig.~\ref{binding_qmc} (c). One can notice that the $-V_\sigma^q$ potentials have
similar density dependence for both quark mass values. This indicates that the slope of
the effective quark mass $m_q^*$ is rather similar, although the $m_q$ values in free
space are different as shown in Fig.~\ref{binding_qmc} (d). In addition, the
$V_\omega^q$ is slightly larger for
$m_q=220$ MeV than that for $m_q =$ 5 MeV, indicating that the light quark with a
larger $m_q$ value would experience more repulsion.
(For a larger value of e.g., $m_q = 430$ MeV, the differences for $m_N^*$ and the light
quark mean field potentials from those of the $m_q = 5$ MeV become more
evident~\cite{deMelo:2018hfw}.)
Using the obtained mean-field potentials for the light (anti)quarks and the effective
(anti)quark masses, we study the meson properties in symmetric nuclear matter.

Recall that the effective quark mass in the nuclear medium is nothing 
but the reflection of the attractive scalar potential, and thus one should not regard 
that it should be positive which applies to the usual physical
mass of the particle ($m_q$ and $m_q^*$ are not observables).

\subsection{Modifications of $\alpha_{s}$ in nuclear medium}

Before addressing the weak-decay constant and distribution amplitude in the nuclear medium, 
we comment on the modifications of $\alpha_s$ in 
Eqs.~(\ref{vcoul})-(\ref{Mmass}) in the medium.  

As reported in Refs.~\cite{Tsushima:1997df,Fuchs:2005zg} (and references therein), 
it is important to satisfy the constraint on the in-medium
$K^+$ total potential that was extracted by heavy-ion simulation and experimental
data, which implies to be 20-30 MeV repulsive at $\rho_0$,
or equivalently, the excitation energy $\omega_{K^+} \equiv M^*_{K^+} +
V^q_\omega$ ($M^*_{K^\pm} = M^*_{K,\bar{K}}$) at
$\rho_0$ to be $\omega_{K^+} = M_{K^+} + (20-30)$ MeV with
$M_{K^+}$ being the free mass.
This constraint should also be satisfied in the present approach.
In the QMC model, $\omega_{K^+}$ can be calculated
directly~\cite{Tsushima:1997df}, but the naive result gives
a small negative value for the $K^+$ total potential at $\rho_0$.
To satisfy the constraint, one may consider
a density dependence of the bag constant $B$~\cite{Lu:1997kw},
and/or the scaling of the kaon vector potential.
However, in the present study,
we consider the medium modifications of $\alpha_s$,
which may also play an important role to reproduce the small positive value of
the $K^+$ total potential at $\rho_0$.
To see the behaviour of $\alpha_s$ in the nuclear medium,
we choose two naive functional forms of the in-medium
$\alpha^*_s$ for this exploratory study,
\begin{equation}
\alpha_s^* \equiv \begin{cases} 
\alpha^*_{s(1)}(\rho) = \alpha_s \left(1 - b_1 \hat{\rho}\right),\\
\\
\alpha^*_{s(2)}(\rho) =\dfrac{\alpha_s}{1 + b_2~{\rm ln}(\hat{\rho}+1)} ,
\end{cases} 
\label{alphas}
\end{equation}
where $\hat{\rho}=\rho/\rho_0$ and dimensionless parameters $b_1 = 0.534$ and $b_2 = 1.650$  are obtained
to satisfy the constraint on $\omega_{K^+}$.
The results for the $\alpha^*_s$ and
$\omega_{K^+} = M^*_{K^+} + V^q_\omega$ are shown in
Fig.~\ref{fig:as_med}. To date, it is not well-known how large the reduction
of $\alpha^*_s$ is in the medium, although it is expected that
$\alpha^*_s$ should decrease as the nuclear density increases, since the Fermi motion
makes the (light) quarks more energetic, and by the QCD's asymptotic freedom.
Using the two $\alpha^*_s$ functional forms, we have
determined the parameters $(b_1,b_2)$ without scaling the vector potential
for the $K^+$ meson (automatically for $K$ and $\bar{K}$ mesons).
If the empirical data are available
for the $K^-$ meson, i.e., $\omega_{K^-} = M^*_{K^-} - V^q_\omega$
($M^*_{K^-} = M^*_{K^+}$), we can deduce whether a scaling for $V^q_\omega$
for $K^\pm$ mesons are necessary or not.
Using the two different density dependencies,
we simulate the modifications of $\alpha^*_s$ in the nuclear medium as shown
in Fig.~\ref{fig:as_med}. The results clearly show that more reduction of
$\alpha^*_s$ results in more repulsion for the total $K^+$ potential ($\omega_{K^+}$).
Recall that the $\alpha_s$ in free space dictates the strength of the attractive
Coulomb-like potential coming from the one-gluon exchange.
We note that $\alpha_s$ is also modified at finite
temperature~\cite{Schneider:2003uz}.

Moreover, in this study, instead of considering the meson loop mechanism
applied for the studies of mass shifts of quarkonia (with no light
quarks) in medium~\cite{Zeminiani:2020aho,Cobos-Martinez:2022fmt},
we apply the in-medium modifications of $\alpha^*_s$ ``effectively'' for all
the quarks treated in this study, according to the quark flavor blindness of QCD.
A more rigorous analysis of the meson mass shift in the medium with more elaborated density-dependent $\alpha_s$ and $\beta$ is left for future studies.

\begin{figure}[t]
	\centering
    \includegraphics[width=0.9\columnwidth]{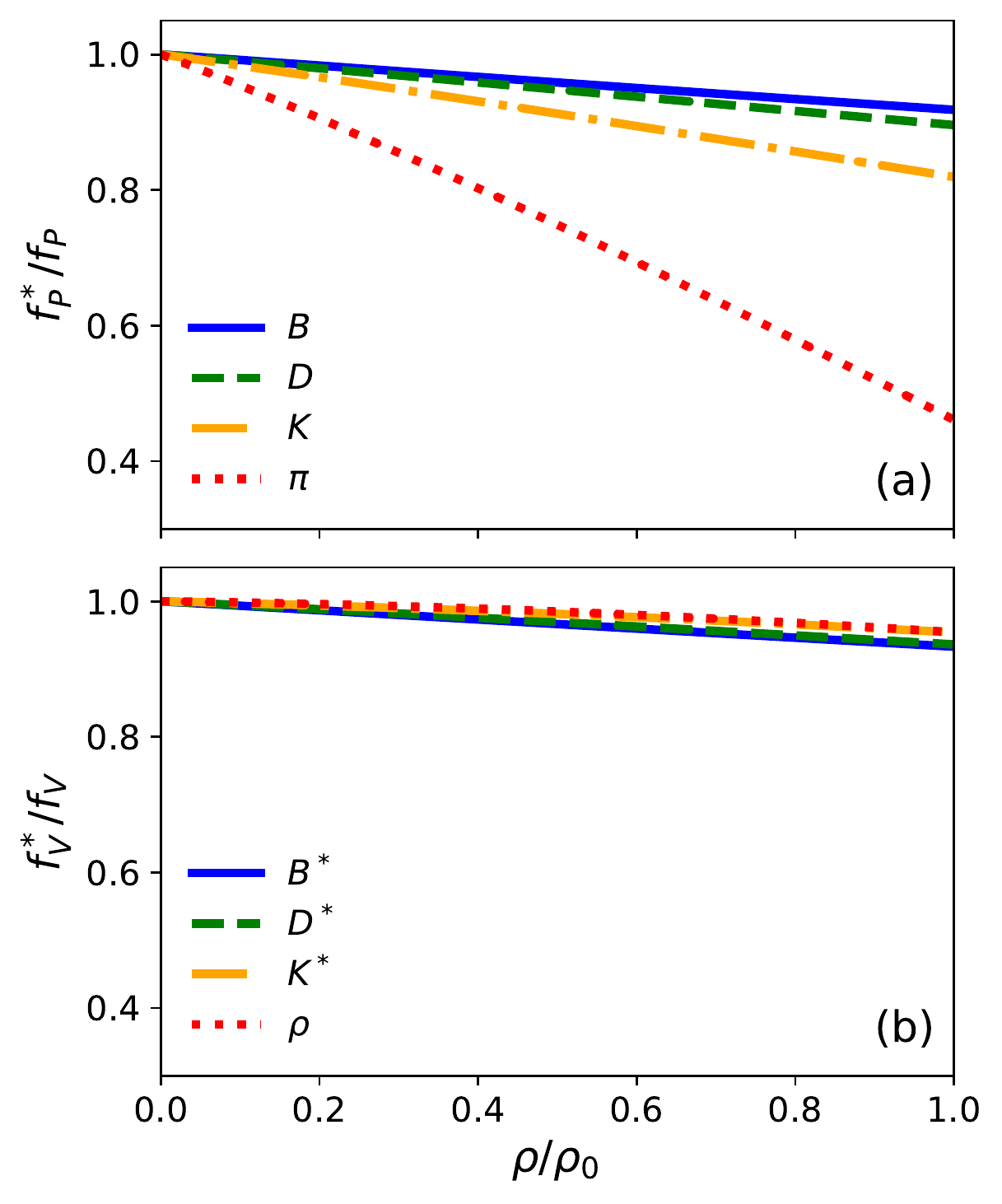}
	\caption{\label{fig:constant} 
The ratios of weak-decay constants in-medium to free space, $f^*_{\rm M}/f_{\rm M}$,
versus
$\rho/\rho_0$ for (a) pseudoscalar and (b) vector mesons. 
For the unequal mass mesons ($q\bar{Q}$ and $Q\bar{q}$), the averaged decay constants 
without 
the vector potential $V_\omega^q$ are presented. 
(See section \ref{sec:medium}.)
The effects of the vector potential are shown in Fig.~\ref{fig:vector}.
}
\end{figure}

\begin{figure}[t]
	\centering
    \includegraphics[width=0.9\columnwidth]{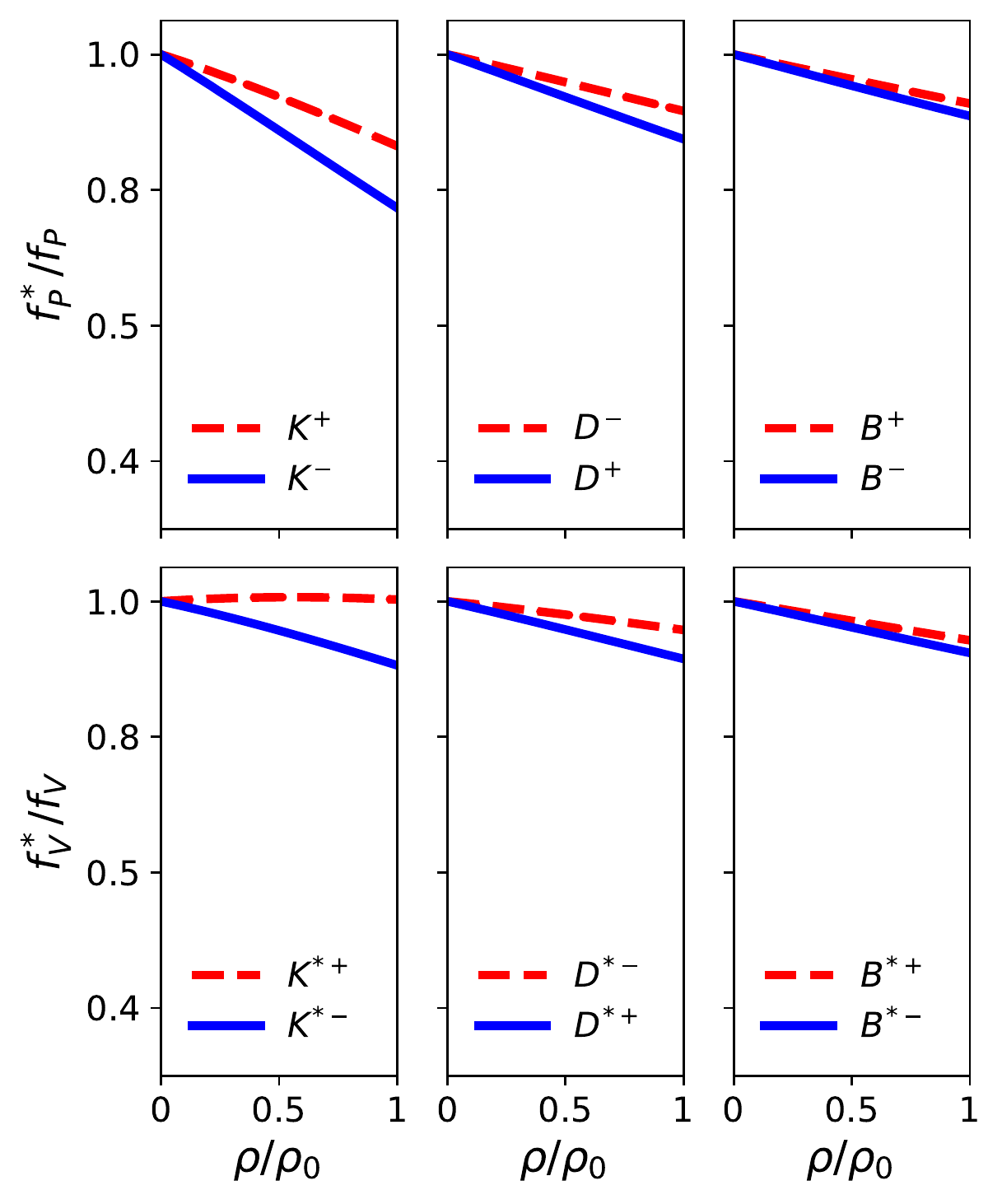}
	\caption{\label{fig:vector} 
Weak-decay constants in symmetric nuclear matter with
the vector potential $V_\omega^q$ 
versus $\rho/\rho_0$ for unequal quark mass mesons
($q\bar{Q}$ and $Q\bar{q}$), $K, D,$ and $B$ mesons.
The impact of $V_\omega^q$ is in a similar order for the pseudoscalar and vector mesons 
corresponding to the quark flavors $s, c$ and $b$ (and their respective antiquarks). 
The effect is more suppressed for heavier flavor mesons.} 
\end{figure}

\subsection{Weak-decay constant in nuclear medium}

Next, we discuss the medium modifications of the averaged weak-decay constants of the light and 
heavy-light pseudoscalar and vector mesons, namely, without the influence of the vector potentials 
for the ratio of $R_{\rm M} = f_{\rm M}^*/f_{\rm M}$ with ${\rm M}={\rm P, V}$ mesons. 
The results are shown in Fig.~\ref{fig:constant}.
On the effects of the vector potentials for the unequal quark mass mesons
($q\bar{Q}$ and $Q\bar{q}$)
such as $K^{(*)}, D^{(*)},$ and $B^{(*)}$,
the results are shown in Fig.~\ref{fig:vector}.

For the lightest pseudoscalar meson, the pion, the ratio of $R_\pi$ without
vector potential
is around 0.5 at $\rho = \rho_0$, showing that the pion weak decay constant in
the medium becomes appreciably reduced relative to that in free space, which is
consistent with that computed in the BSA
calculation with the mass regulator~\cite{deMelo:2014gea,deMelo:2016uwj}.
In comparison with the deeply bound pionic atom experiment, our result is smaller than the 
empirical value $R_\pi^2\approx 0.64$~\cite{Suzuki:2002ae}. We also notice that the
reduction is sensitive to the quark mass value used.
We separately check for the larger 
$m_q$ value and the less reduction is obtained, which is consistent with the
the BSE-NJL model results in larger quark
mass~\cite{Hutauruk:2018qku,Hutauruk:2019ipp}.
However, overall, our predicted results are consistent with other theoretical calculations, 
namely, the weak-decay constant decreases as the nuclear density increases.

While results for the $R_K$, $R_D,$ and $R_B$ without vector potential are
found to have smaller reductions, which are
approximately about $0.8 -0.9$ at $\rho = \rho_0$, as shown in Fig.~\ref{fig:constant} (a). 
Also, it shows a large reduction of $R_\pi$, compared 
to the other pseudoscalar mesons. This can be understood by the fact that the pion contains 
purely light quarks and antiquark, and the two light quarks are subject to feel
stronger medium effects since the light quark chiral condensates reduce
faster as the nuclear density increases than those of the heavier quarks.
In contrast, the quark condensates for the $s$, $c$, and $b$ quarks change very slowly 
and a small amounts in the nuclear medium. (See again the discussions made in the 
introduction part of Ref.~\cite{Tsushima:2020gun}.) 
The in-medium weak decay constant ratios at $\rho_0$ follow the order as
\begin{eqnarray}
1 > \frac{f_B^*}{f_B} > \frac{f_D^*}{f_D} > \frac{f_K^*}{f_K} > \frac{f_\pi^*}{f_\pi}.
\end{eqnarray}
For the heavy-light mesons, again it is clearly shown that the reductions of the weak-decay 
constants are less pronounced. 
This is because the medium effect becomes smaller relative to the 
masses of the heavy quarks and they weakly couple in the medium.
(In the present approach, mean fields do not couple directly to the heavy
quarks at the lowest order.)

Now, we discuss the crucial role of the vector potential ($\pm V_\omega^q$ for $q$ and 
$\bar{q}$) on the in-medium weak-decay constants for the positively and negatively charged states  
for $K^{\pm}, D^{\mp}$ and $B^{\pm}$. Figure~\ref{fig:vector} demonstrates that the effect of 
$V_\omega^q$ on the ratios of the in-medium weak-decay constants obviously depends on 
the light quark contents that are controlled by the term of $(1 \pm V_\omega^q/M_0^*)$. 
It shows that the differences in the weak-decay constant ratios for the 
heavy-light mesons with opposite charge states are getting suppressed 
as the mesons become heavier, since the invariant meson mass $M_0^*$ 
appearing in the denominator becomes larger, and thus the effect of the vector potential 
is relatively suppressed by $M_0^*$. 
Therefore, the effect of $V_\omega^q$ is more pronounced for the lighter mesons,   
and suppressed for the heavier mesons.

For the light and heavy-light vector mesons, the ratios of the in-medium weak decay constants 
are $R_{\rm V} \simeq 0.95$ at $\rho_0$ without the vector potential.
The order of the ratios for the in-medium vector meson decay constants have opposite order 
to those for the pseudoscalar mesons:
\begin{eqnarray}
\frac{f_{B^*}^*}{f_{B^*}}  < \frac{f_{D^*}^*}{f_{D^*}} < \frac{f_{K^*}^*}{f_{K^*}} 
< \frac{f_\rho^*}{f_\rho} < 1.
\end{eqnarray}
Also, it is found that the differences among the ratios of the different vector mesons are rather 
small as can be seen in Fig.~\ref{fig:constant} (b).
Moreover, the effect of the $\pm V_\omega^q$ for the in-medium 
vector meson decay constants have a similar tendency to that 
for the pseudoscalar mesons, as shown in Fig.~\ref{fig:vector}.
However, the magnitude of the ratios is different from that for pseudoscalar 
mesons, in particular for $K^{*\pm}$.

The different behavior of $R_{\rm M}$ between the pseudoscalar and vector mesons can be 
qualitatively explained by the weak-decay constant formula. 
For instance, the operators in Eqs.~(\ref{eq:operator1}) and 
(\ref{eq:operator2}) for $\pi$ and $\rho$ mesons, which are respectively given by 
\begin{eqnarray}
\mathcal{O}_{\rm P}^* &=& m_q^*,\\
\mathcal{O}_{\rm V}^* &=& m_q^*  + \frac{2\mathbf{k}_\perp^2}{M_0^* + 2m_q^* }.
\label{OV}
\end{eqnarray}
If we assume SU(6) symmetry, the difference between the in-medium decay constant 
of $\pi$ and $\rho$ mesons comes from the second term of $\mathcal{O}_{\rm V}^*$.
Since $m_q^*$ decreases in the medium as density increases, 
the first term of $\mathcal{O}_{\rm V}^{*(1)}$ decreases, 
while the second term $\mathcal{O}_{\rm V}^{*(2)}$ increases as density increases. 
For this reason, the decay constant for the $\pi$ meson is significantly
reduced in the medium.

\begin{figure}[t]
	\centering
	\includegraphics[width=0.98\columnwidth]{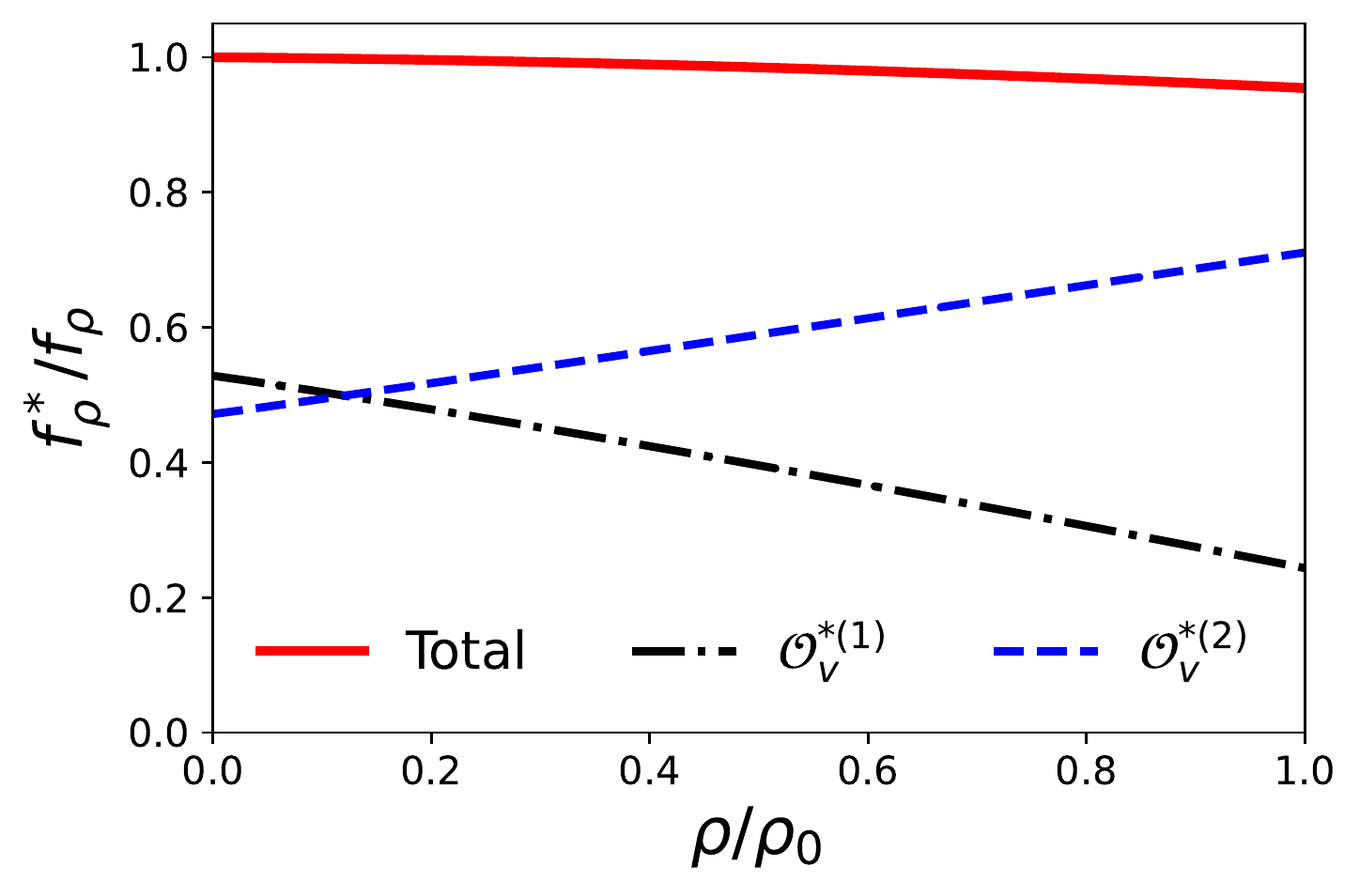}
	\caption{\label{fig:rho} 
Contributions of the first term $\mathcal{O}_{\rm V}^{*(1)}=m_q^*$ and the second term 
$\mathcal{O}_{\rm V}^{*(2)}= 
2\mathbf{k}_\perp^2/(M_0^* + 2m_q^*)$ of the operator $\mathcal{O}_{\rm V}^*$ 
for the weak-decay constant of $\rho$ meson in medium versus $\rho/\rho_0$.}
\end{figure}

The results for the weak-decay constants of $\rho$ meson as well as the other
vector mesons are found to be almost constant and nearly without the influence
of medium effect, due to the competing contributions between the two terms for
the total vector-meson decay constants as shown in Fig.~\ref{fig:rho}.
One can see that the second term $\mathcal{O}_{\rm V}^{*(2)}$ in Eq.~(\ref{OV})
dominates at $\rho_0$, while the first and second terms contribute rather
similarly to the decay constant in free space. For heavy-light vector mesons,
the contribution from $\mathcal{O}_{\rm V}^{*(2)}$ is
rather suppressed even in free space due to heavy-flavor quark mass.
As a result, the medium modifications of $f_{\rm P}^*/f_{\rm P}$ and $f_{\rm V}^*/f_{\rm V}$ 
for the heavy mesons are rather similar as shown
in Figs.~\ref{fig:constant} (a) and (b).

\begin{figure}[t]
	\centering
	\includegraphics[width=0.98\columnwidth]{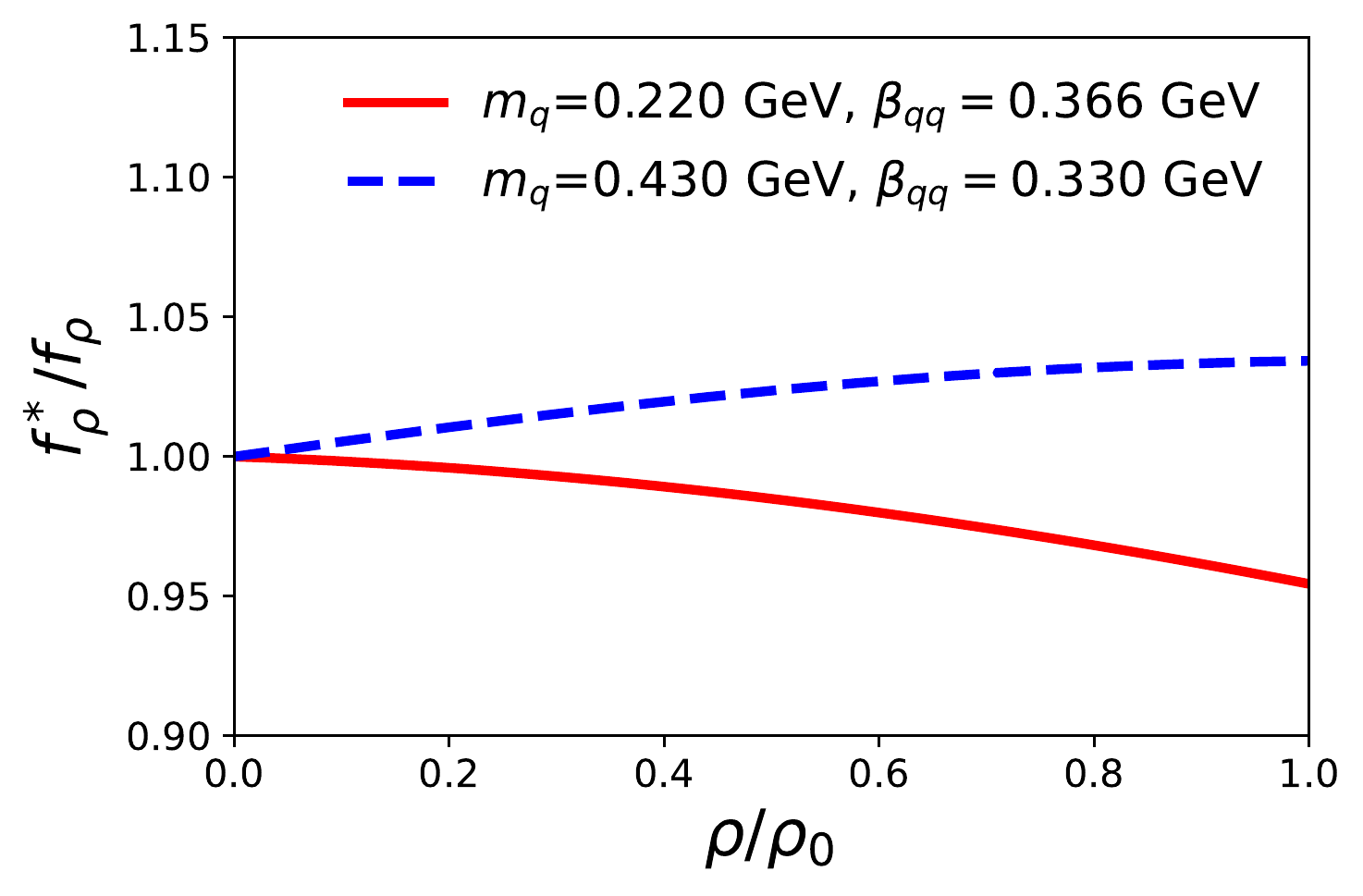}
	\caption{\label{fig:compare} 
	Ratios of $f_\rho^*/f_\rho$ for different quark mass values versus
$\rho/\rho_0$. The Gaussian
parameter $\beta_{q\bar{q}}$ is fixed to reproduce the decay constant $f_\rho^{\rm expt}$, 
when we use $m_q=$ 0.430 GeV.}
\end{figure}
\begin{figure}[t]
	\centering
	\includegraphics[width=0.98\columnwidth]{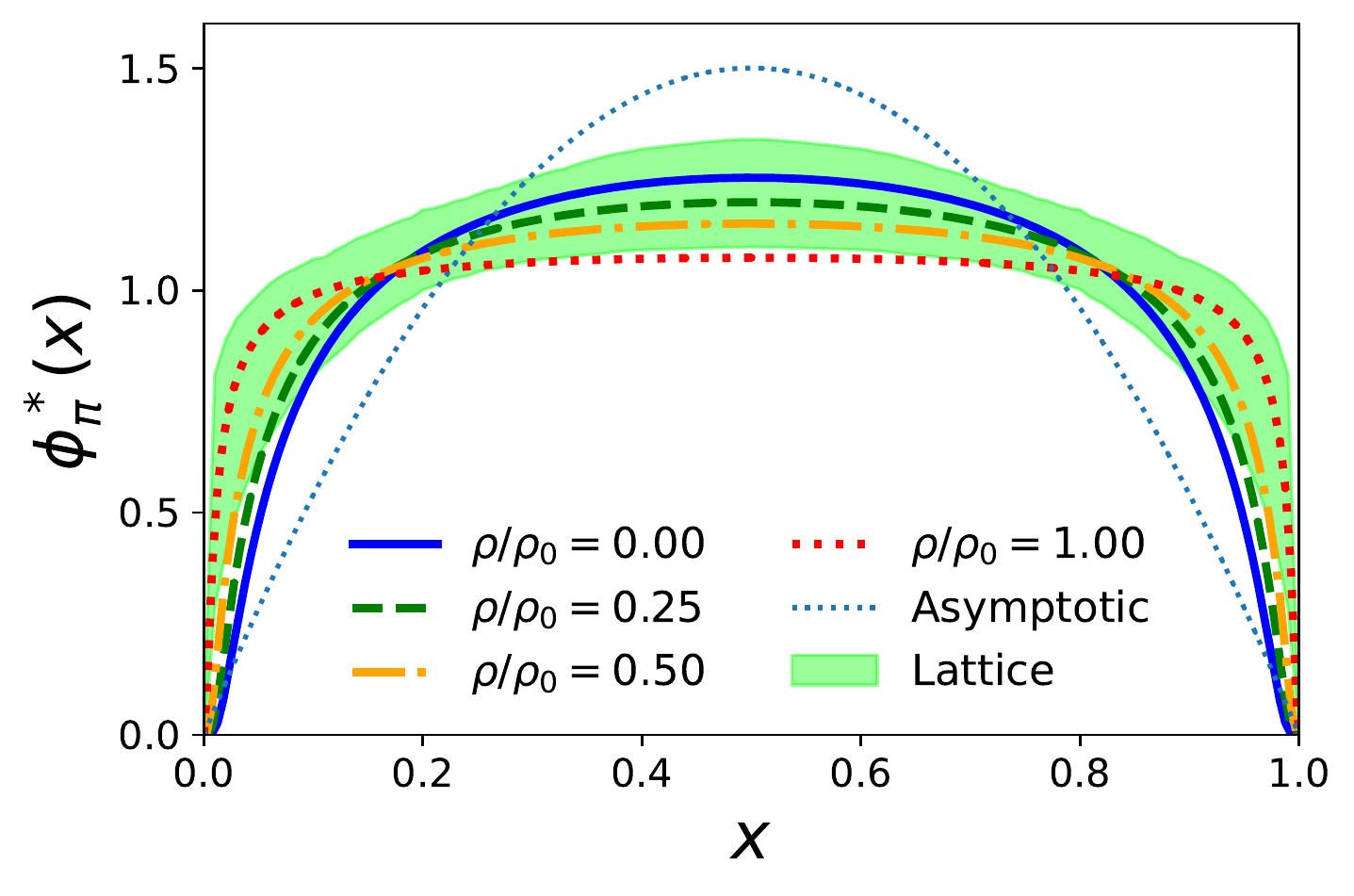}
	\includegraphics[width=0.98\columnwidth]{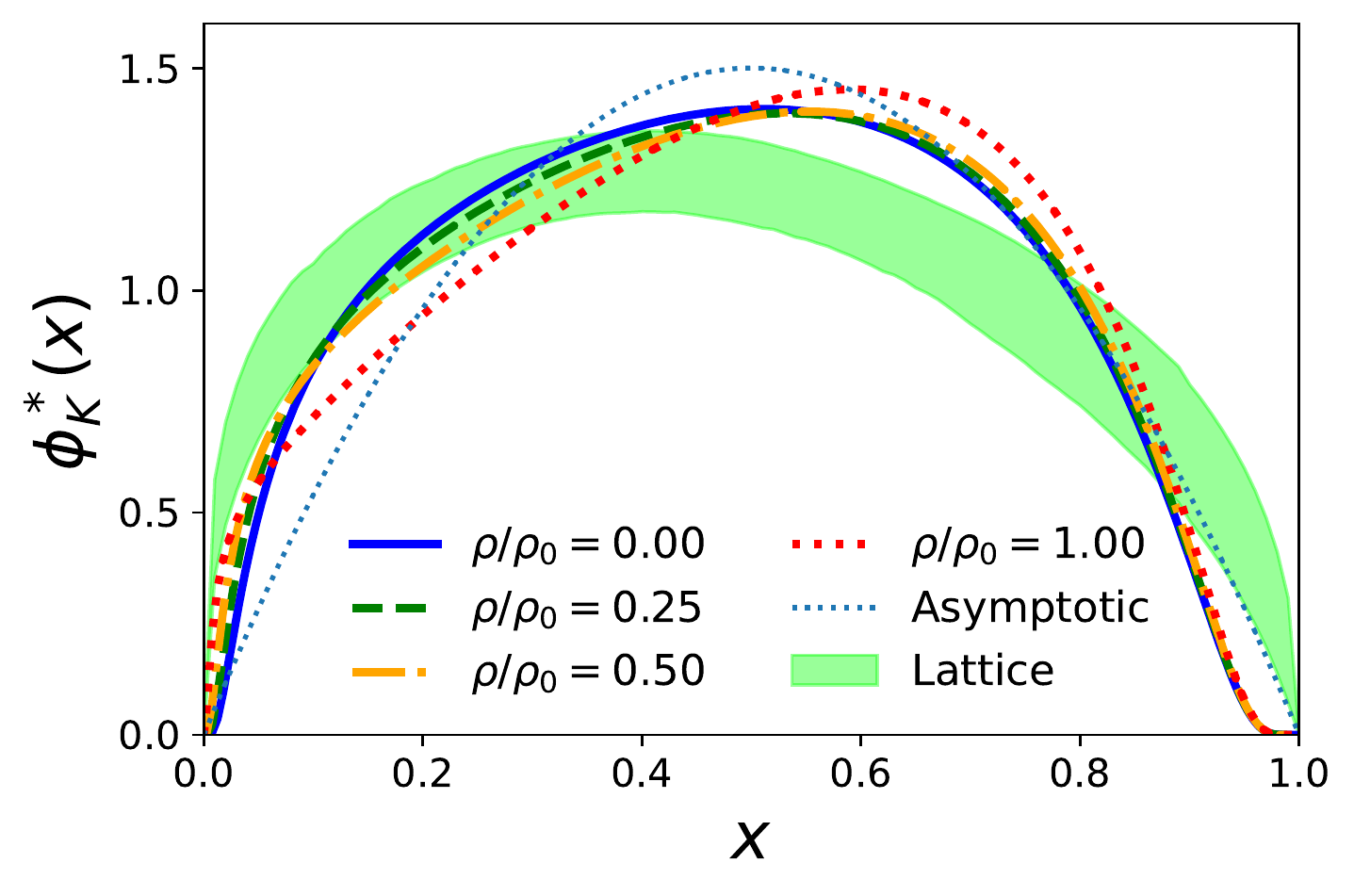}
	\caption{\label{fig:da_pion} 
	In-medium DAs of the pion (upper panel) and kaon (lower panel) with several densities 
versus longitudinal momentum $x$. The lattice QCD data in free space is
represented by a green-shaded region, which is taken from
Ref.~\cite{LatticeParton:2022zqc}.
The predicted results for the DAs in the asymptotic region are represented by the smallest dotted 
lines.} 
\end{figure}

In the following, we study the effect of the different quark mass values
$m_q$ on the weak-decay constant $f_\rho^*$.
To do so, we use $m_q = 0.430$ GeV so that we can compare our
result with that obtained in Ref.~\cite{deMelo:2018hfw},
although they calculated in the light-front
constituent quark model based on the BSA approach with a regulator mass,
while, in the present work, we employ the Gaussian parameter with
$\beta_{q\bar{q}}=$ 0.330 GeV to reproduce the empirical value $f_\rho^{\rm
expt.} = 269$ MeV.
Results with the two values of $m_q$, 0.220 and 0.430 GeV are shown in
Fig.~\ref{fig:compare}. 
The results of the $f_{\rho}^*/f_\rho$ ratio with $m_q =$ 0.220 GeV smoothly decrease as density increases, 
while the result with $m_q =$ 0.430 GeV slightly
increases as the density increases.

Similar behavior of the weak-decay constants can be found for the other vector
mesons in our approach.
The increase of $f_\rho^*$ in medium  with $m_q = 0.430$ GeV is also observed in 
Ref.~\cite{deMelo:2018hfw}, however, it shows an oscillating behavior as nuclear density increases. 
Note that, the increasing rate is smooth in the present approach, as seen in 
Fig.~\ref{fig:compare}. The differences in the $f_\rho^*$ increasing behavior
with that of Ref.~\cite{deMelo:2018hfw} may be attributed to the use of the
invariant mass $M_0$ in the present approach.
Overall, our result for the in-medium $\rho$ meson decay constant 
$f_\rho^*$ with $m_q =$ 0.430 GeV is consistent with
Ref.~\cite{deMelo:2018hfw}.
We argue that the moderate increase of $f_\rho^*$ in the present approach  
can be attributed to the use of the invariant mass $M_0$ in 
calculating the weak-decay constant, which is a special aspect of the present
LFQM, where it was shown in Ref.~\cite{Arifi:2022qnd} that the use
of the invariant mass $M_0$ enables to calculate the $\rho$-meson weak-decay
constant from various light-front current components and polarizations,
and it can produce the weak-decay constant self-consistently.

\subsection{Distribution amplitude in nuclear medium}

\begin{figure}[t]
	\centering
	\includegraphics[width=0.98\columnwidth]{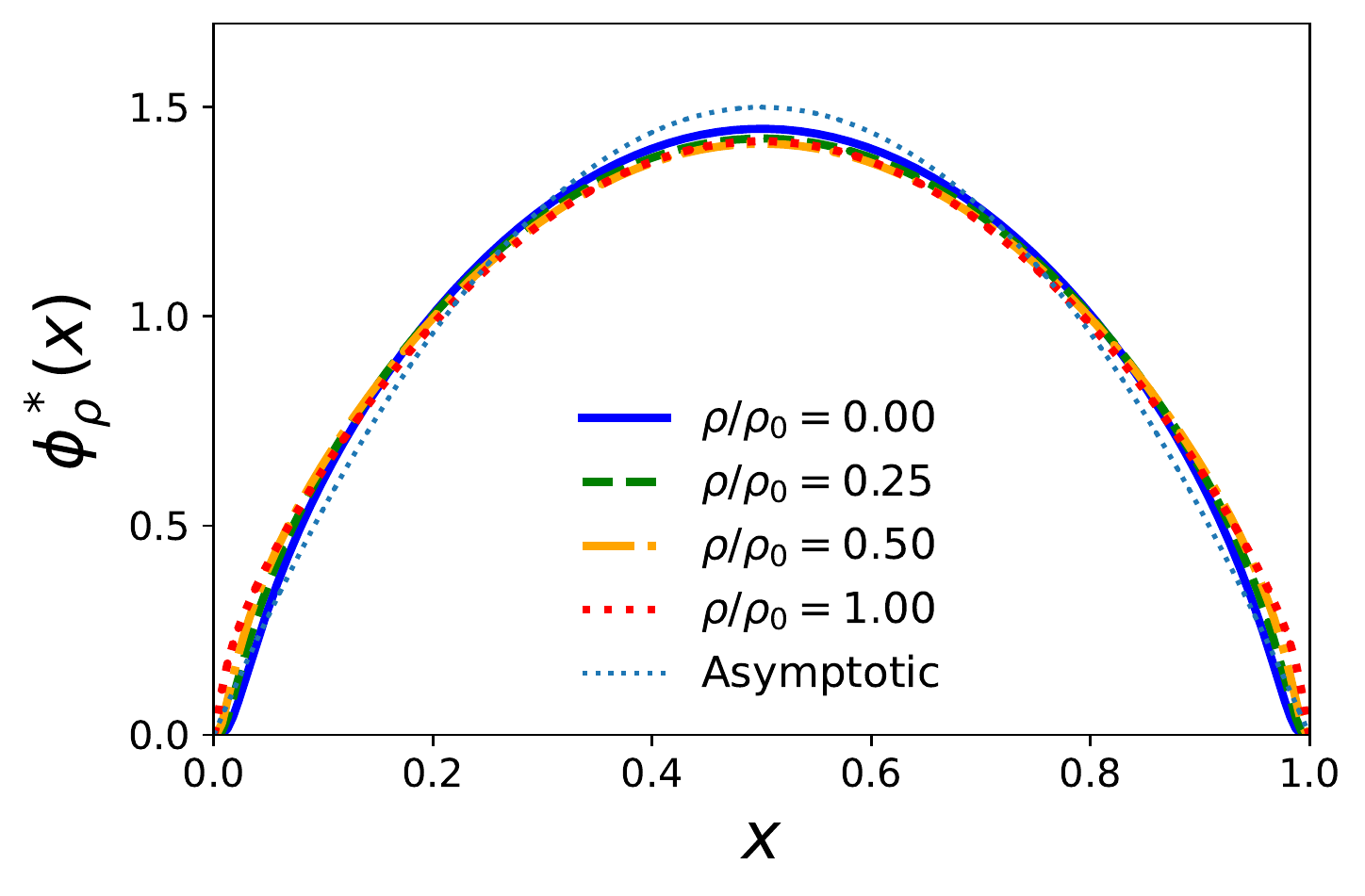}
	\includegraphics[width=0.98\columnwidth]{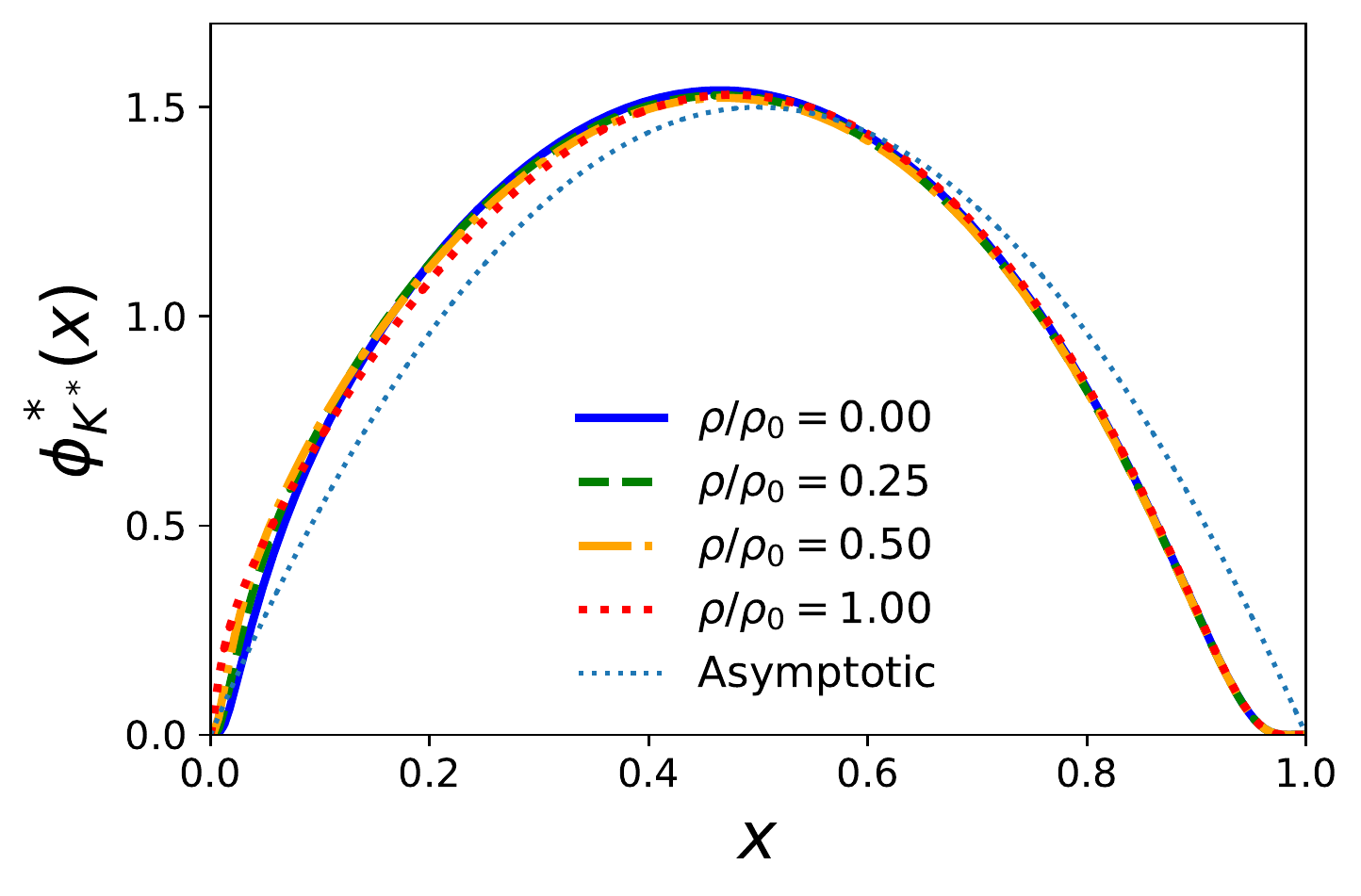}
	\caption{\label{fig:da_rho} 
	In-medium DAs of the $\rho$ (upper panel) and $K^*$ (lower panel) vector mesons versus $x$ 
for several densities. See also the caption of Fig.~\ref{fig:da_pion}.
} 
\end{figure}

\begin{figure*}[t]
	\centering
    \includegraphics[width=1.9\columnwidth]{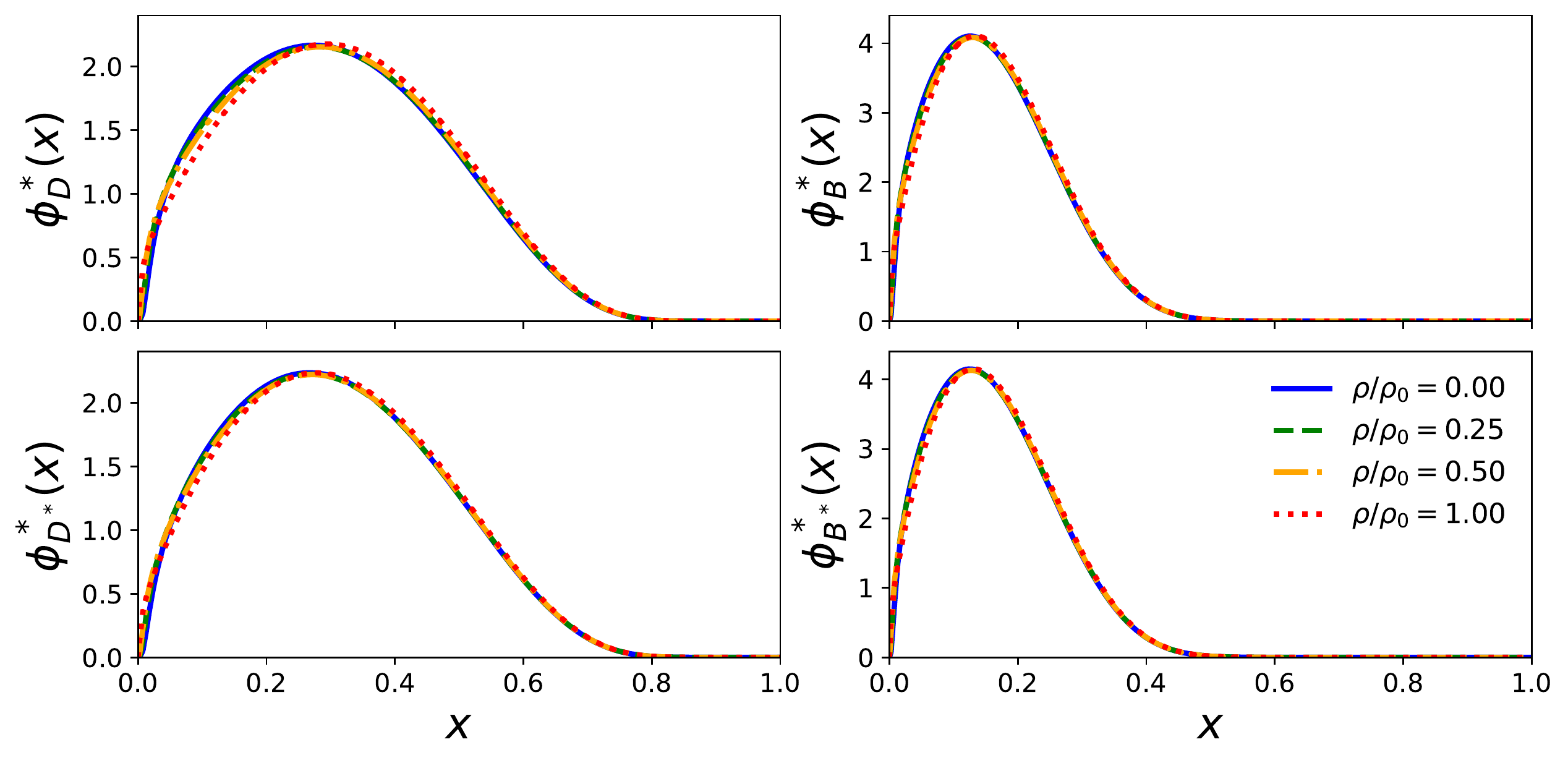}
	\caption{\label{fig:da_heavy} 
	Same as in Fig.~\ref{fig:da_rho}, but for the DAs of $D^{(*)}$  and $B^{(*)}$ in medium.} 
\end{figure*}

Here, we present our final results for the in-medium DAs of the light and heavy-light pseudoscalar 
and vector mesons. We set the longitudinal momentum $x$ carried by the
light quark $q$ in $(q\Bar{Q})$ meson and by antiquark $\Bar{q}$ in
$(Q\Bar{q})$ meson.
Furthermore, for simplicity, we present the average, or equivalently, the
results calculated without the vector potential $V_\omega^q$. The in-medium DAs
for the pion and kaon are presented in
Fig.~\ref{fig:da_pion}, while those for the $\rho$ and $K^*$ are shown in 
Fig.~\ref{fig:da_rho}, and those for the heavy-light pseudoscalar and vector
mesons are depicted in Fig.~\ref{fig:da_heavy}.

In Fig.~\ref{fig:da_pion}, the calculated DAs for the pion (upper panel) and
kaon (lower panel) are compared with the predicted asymptotic ones (dotted
line) as well as the recent lattice
QCD data (green shaded region)~\cite{LatticeParton:2022zqc}. Very recently, another lattice QCD 
result for pion DA is reported ~\cite{Holligan:2023rex} and they find that
their result lies within the green shaded
region that obtained in Ref.~\cite{LatticeParton:2022zqc}.
We observe that the pion DA in free space~\cite{Choi:2007yu} is
consistent with the recent lattice QCD results.
In the medium, the pion DA becomes flatter with the value approaching unity 
at $\rho = \rho_0$, but the $x$ dependence of DA becomes broadened.
This is expected due to the reduction of the quark mass $m_q^*$ in the medium, 
reflecting the effect of the partial restoration of the chiral symmetry.

Results for the kaon DA in free space as well as in a medium are shown
in the lower panel of Fig.~\ref{fig:da_pion}, where the result in free space is
calculated with the parameters of Ref.~\cite{Choi:2007yu}, 
and has a reasonable agreement with the lattice QCD data especially in the $x<0.5$ region.
However, the kaon DA in free space overestimates the lattice data in the $0.5 <x <0.8$ region and underestimates the asymptotic prediction 
in the small $x$ region. 
Compared with the lattice QCD data, the kaon DA is
consistent with lattice data in the near endpoint region at $x \to 0$.
However, at the near-end point $x\to$ 1, the result is suppressed and shows
different behavior from the lattice QCD data.
It is worth noting that the fast suppression of the Kaon DA near the endpoint $x \to$ 1 may indicate that the kaon DA is sensitive to the choice of the radial wave function.
In Ref.~\cite{Choi:2017uos}, the authors compared the twist-3 DA of pion and kaon with two different trial wave functions, 
i.e., the Gaussian and Power-law radial wave function. 
One of the apparent differences is that the DA near-end point with the Power-law wave function is more enhanced as compared with the Gaussian one.
Recall that the Gaussian radial wave function is employed in the present
work. 
Furthermore, the DA near endpoints cannot be computed by the lattice QCD directly, the values are obtained from a phenomenological extrapolation that may give some ambiguity~\cite{LatticeParton:2022zqc}.
Another possible source of discrepancy is the SU(3) flavor symmetry-breaking effect. In this work, the $(m_q,m_s)=(220,450)$ MeV is taken which shows a considerable difference. 
A smaller quark mass difference may be favored as far as the lattice data are concerned. 
Further analysis of the kaon DA in free space is needed by modifying the trial wave function to get the best fit to the lattice QCD.
As for the kaon DA in the medium, it is found to have the largest reduction at around $x=$ 0.2, whose peak slightly moves to the larger $x$ region, since the (effective) mass difference between the light quark and the strange quark in the medium is enhanced.

Results for the $\rho$ and $K^*$ DAs are shown in the upper and lower panels of 
Fig.~\ref{fig:da_rho}, respectively. 
The DA for the $\rho$ is consistent with that for the asymptotic result. 
The $\rho$ DA is moderately modified in the medium. 
However, the $K^*$ DA shows different behavior from the asymptotic results. 
It is shifted to the smaller $x$ region, which 
is expected due to the SU(3) flavor symmetry breaking in the quark masses. 
Similar to the $\rho$ DA, the $K^*$ DA is moderately modified in the medium, as shown in the lower 
panel of Fig.~\ref{fig:da_rho}. 
Note that the small-medium modifications of the vector meson DAs 
are in conformity with the small reduction of $f_{\rm V}^*$ in medium, 
as shown in Fig.~\ref{fig:constant}.

Our results for the heavy-light meson $B, D, D^*$, and $B^*$ DAs are shown
in the left and right panels of Fig.~\ref{fig:da_heavy}.
The results indicate that the heavy-light meson DAs are nearly unmodified in the medium. 
A similar reason for the kaon with the strange quark content, 
can be understood by the heavy quark contents of the
heavy-light mesons, where the heavy quarks are not modified directly in the
medium.

\begin{figure*}[t]
	\centering
	\includegraphics[width=1.9\columnwidth]{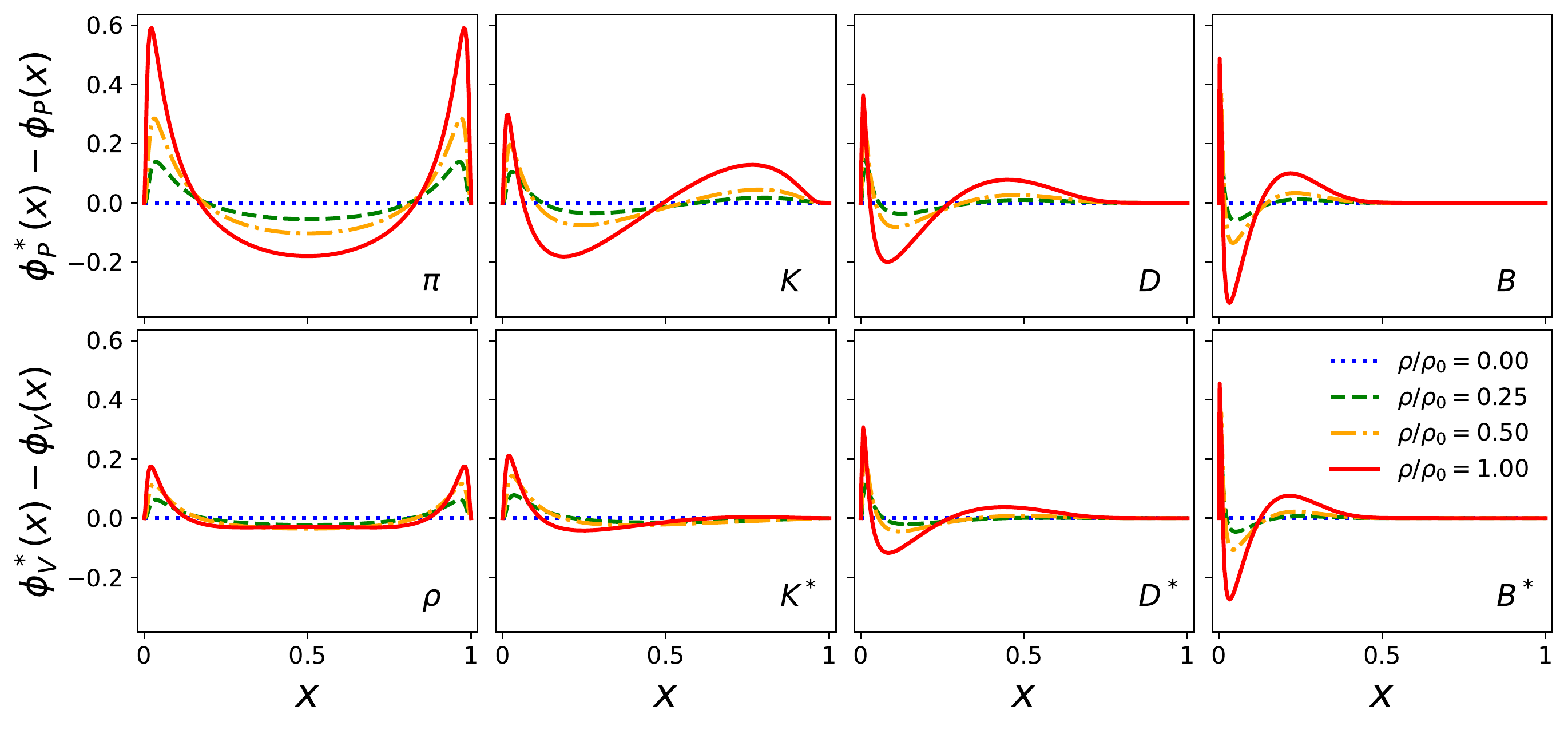}
	\caption{\label{fig:da_diff} 
	The difference of DAs defined by $\Delta\phi_M(x) = \phi_M^*(x) -  \phi_M(x) $ 
	for various mesons.} 
\end{figure*}

Since the smaller in-medium DA modifications for the light and
heavy-light mesons cannot easily be seen
from the results shown in Figs.~\ref{fig:da_rho} and \ref{fig:da_heavy},
respectively, we show the difference between the DAs in the
medium [$\phi_{\rm M}^*(x)$] and in free space [$\phi_{\rm M}(x)$], defined by
\begin{eqnarray}
\Delta\phi_M(x) = \phi_{\rm M}^*(x) - \phi_{\rm M}(x).
\end{eqnarray}
The DA differences in free space and in the medium are shown in Fig.~\ref{fig:da_diff}. 
The effects of the in-medium modifications of DAs are more evident for each
meson. The results show that the effects of the in-medium modifications are not
negligibly small for the heavy-light mesons $B, D, D^*$, and $B^*$, in
particular, at small $x$.
However, the medium effect for the $B$ and $B^*$ mesons is almost
negligible at around $x > 0.5$, and the medium effect for 
the $D$ and $D^*$ mesons is also negligible at around $x > 0.8$. 
Figure~\ref{fig:da_diff} clearly shows that the medium effect on the DAs  for
the light pseudoscalar meson is more pronounced compared with that for vector
mesons. Also, the medium modifications of DAs for the light pseudoscalar mesons
are similarly larger than those for the weak-decay constants compared with the
other mesons.

\section{Summary} \label{sec:summary}
To summarize, in the present work, we have investigated the in-medium
modifications of the weak-decay constants and distribution amplitudes (DAs) of
the light and heavy-light pseudoscalar and vector mesons. For this purpose, we
have constructed a combined model based on a light-front quark model (LFQM)
that describes the meson properties both in free space
as well as in a nuclear medium based on an equal footing, and the
quark-meson coupling (QMC) model to simulate the medium effects on the
(light) quarks. The in-medium quark properties are modified by the
self-consistent scalar and vector mean fields
generated by the surrounding nucleons.
It is evident that the light quark mass is reduced in the
medium, reflecting that the chiral symmetry is partially restored. 
We note that the present approach is similar to that was practiced in 
Refs.~\cite{deMelo:2016uwj,deMelo:2014gea,deMelo:2018hfw}, 
but we use a different approach, a more QCD dynamics-motivated light-front quark
model with the Gaussian wave function.

We have found that the weak decay constants for the pseudoscalar mesons  with
light-light quark contents decrease relatively faster in the medium as nuclear
density increases compared with those for the vector mesons with the
light-light and heavy-light quark contents
(as in Fig.~\ref{fig:constant}). 
For the less pronounced in-medium modifications of the vector meson 
decay constants are, due to the competition between the decreasing and
increasing contributions of the $\mathcal{O}_{\rm V}^{*(1)}$ and
$\mathcal{O}_{\rm V}^{*(2)}$ operators in the medium.
We should note that the medium modifications of 
weak-decay constants are sensitive to the choice of the free space quark mass value.
The use of the larger values of the quark mass value leads to a less reduction of
the weak-decay constant.

The role of the vector potential $V_\omega^q$ is also studied for the mesons with
heavy-light mesons (as in Fig.~\ref{fig:vector}). We have found that the vector
potential modifies differently the decay constants within the meson multiplet members,
namely for the different charges (isospin) cases, by a factor $(1\pm V_\omega^q/M_0^*)$.
Also, we have found that the trend for the effect of the vector potential
is similar for the pseudoscalar and vector mesons, but the effect is much more
suppressed for the heavy mesons due to the factor by the invariant mass, $1/M_0^*$.

Interestingly, we have found that the pion and kaon DAs are more modified
in medium among all the mesons studied in the present study.
The medium modifications of DAs can be clearly visualized by calculating the DA
differences in free space and in a medium, namely by calculating
$\Delta \phi_{\rm M} (x) =\phi_{\rm M}^*(x)-\phi_{\rm M} (x)$ (as in
Fig.~\ref{fig:da_diff}).
It is found that the shape of DAs in medium has large enhancement in the
endpoint regions, while it is reduced in the moderate $x$ for the light mesons. 
The enhancement near the endpoint at $x\to 1$ is smeared and shifted to
the lower $x$ region.
 
We expect that the present results may provide useful guidance for constructing more
sophisticated models based on the quark degrees of freedom to investigate
the in-medium properties of the light
and heavy-light pseudoscalar and vector mesons. 
Also, the present study will provide useful information on the
possible meson-nuclear bound states, that are planned to measure in the modern
experiment
facilities~\cite{n-PRiMESuper-FRS:2016vbn,Itahashi:2012ut,WASA-at-COSY:2020bch}.
Furthermore, the experiment via the pionic or kaonic Drell-Yan reaction process 
with the heavy nuclear targets can be a way to probe the modifications of DAs 
in nuclear medium and nuclei~\cite{Chang:2013opa}. 
In principle, our present approach can be applied to investigate various
meson properties in medium, for instance, in-medium meson electromagnetic
form factors. Such studies are underway.

\vspace{1cm}
\section*{Acknowledgements}
P.T.P.H. thanks Jian-Hui Zhang (Lattice Parton Collaboration) for providing us
with the recent lattice QCD data for the kaon and pion distribution amplitudes.
A.J.A. was supported by Special Postdoctoral Researcher (SPDR) Program at RIKEN  and the Young Scientist Training (YST) Program at the Asia Pacific
Center for Theoretical Physics (APCTP) through the Science and Technology Promotion
Fund and Lottery Fund of the Korean Government and also by the
Korean Local Governments-Gyeongsangbuk-do Province and Pohang City.
The work of P.T.P.H. was supported by the National Research Foundation of Korea (NRF)
grants funded by the Korean government (MSIT) Nos. 2018R1A5A1025563, 2022R1A2C1003964,
and 2022K2A9A1A0609176.
The work of K.T. was supported by
the Conselho Nacional de Desenvolvimento Cient\'{i}fico
e Tecnol\'{o}gico (CNPq) Process, No.~426150/2018-0, and No.~304199/2022-2, and
Funda\c{c}\~{a}o de Amparo \`{a} Pesquisa do Estado de S\~{a}o Paulo (FAPESP) Process,
No.~2019/00763-0, and the work was also part of the
projects, Instituto Nacional de Ci\^{e}ncia e Tecnologia --- Nuclear Physics and
Applications (INCT-FNA), Brazil, Process. No.~464898/2014-5.

\vspace{3ex}
\noindent

\end{document}